\documentclass[lettersize,journal]{IEEEtran}
\usepackage[switch]{lineno} % line number
\usepackage{amsmath,amssymb,amsfonts,amsthm}
\usepackage{algorithmic}
\usepackage{algorithm}
\usepackage{array}
\usepackage[caption=false,labelfont=sf,textfont=sf]{subfig}
\usepackage{textcomp}
\usepackage{stfloats}
\usepackage{url}
\usepackage{verbatim}
\usepackage{graphicx}
\usepackage{cite}

\usepackage{xcolor}
\usepackage{booktabs}
\usepackage{multirow}
\usepackage{mathtools}
\usepackage{mathrsfs}
\usepackage{enumitem}
\begin{document}

\title{DeepCRF: Deep Learning-Enhanced CSI-Based RF Fingerprinting for Channel-Resilient WiFi Device Identification
}

\author{Ruiqi~Kong and
        He~(Henry)~Chen
\thanks{The authors are with the Department of Information Engineering, The Chinese University of Hong Kong, Hong Kong SAR, China. H. Chen is also with Shun Hing Institute of Advanced Engineering, The Chinese University of Hong Kong, Hong Kong SAR, China. (E-mail: \{kr020, he.chen\}@ie.cuhk.edu.hk)}\\ 
\thanks{Part of the work has been presented in 
2024 IEEE Conference on Computer Communications Workshops (INFOCOM WKSHPS) \cite{kong2024channelresilient}.}
}

\maketitle

\begin{abstract}
This paper presents DeepCRF, a new framework that harnesses deep learning to extract subtle micro-signals from channel state information (CSI) measurements, enabling robust and resilient radio-frequency fingerprinting (RFF) of commercial-off-the-shelf (COTS) WiFi devices across diverse channel conditions. Building on our previous research, which demonstrated that micro-signals in CSI, termed micro-CSI, most likely originate from RF circuitry imperfections and can serve as unique RF fingerprints, we develop a new approach to overcome the limitations of our prior signal space-based method. While the signal space-based method is effective in strong line-of-sight (LoS) conditions, we show that it struggles with the complexities of non-line-of-sight (NLoS) scenarios, compromising the robustness of CSI-based RFF. To address this challenge, DeepCRF incorporates a carefully trained convolutional neural network (CNN) with model-inspired data augmentation, supervised contrastive learning, and decision fusion techniques, enhancing its generalization capabilities across unseen channel conditions and resilience against noise. Our evaluations demonstrate that DeepCRF significantly improves device identification accuracy across diverse channels, outperforming both the signal space-based baseline and state-of-the-art neural network-based benchmarks. Notably, it achieves an average identification accuracy of 99.53\% among 19 COTS WiFi network interface cards in real-world unseen scenarios using 4 CSI measurements per identification procedure.
\end{abstract}

\begin{IEEEkeywords}
Radio-frequency fingerprinting, channel state information, micro-CSI, WiFi, deep learning.
\end{IEEEkeywords}

\section{Introduction} 
\IEEEPARstart{T}{he} rapid expansion of low-cost wireless devices under the Internet of Things (IoT) paradigm, coupled with the broadcast nature of wireless transmission mediums, poses significant security challenges for wireless networks \cite{jagannath2022comprehensive}. Traditional security and network management systems expect each physical node to have a unique identity, following a one-to-one node-to-identity mapping. However, this assumption is compromised by practices such as MAC address randomization and vulnerabilities like Sybil attacks \cite{abbas2021improving}. These issues allow multiple identities to originate from a single device, which can appear as multiple devices, complicating network management and anomaly detection, and leading to resource exhaustion from illegitimate use. Therefore, it is crucial to develop a consistent and resource-efficient identifier for uniquely identifying transmitters.

Radio-frequency fingerprinting (RFF) emerges as a promising solution, which leverages the distinctive and inevitable manufacturing imperfections present in the RF components of wireless devices as identifiers/fingerprints. Such hardware imperfections lead to subtle signal distortions that, while typically having a negligible effect on the decoding of transmitted data, provide a sufficiently distinctive fingerprint for each device. These fingerprints are remarkably consistent and difficult to mimic or tamper with \cite{zhang}. Such characteristics of RFF prove valuable in mitigating identity threats by ensuring that each physical device can be associated with a single, consistent identity \cite{abbas2021improving}. 
The promise of RFF systems has been substantiated through its capability of identifying wireless devices transmitting identical information or produced by the same manufacturer \cite{xu2015device,jagannath2022comprehensive}. Furthermore, RFF systems offer a significant advantage in terms of computational efficiency. Unlike cryptographic methods that demand considerable resources for key generation and management, RFF seizes the intrinsic characteristics of RF hardware as a natural identifier \cite{9450821}. This not only provides a seamless security layer but also aligns with the resource-conscious ethos of the IoT ecosystem. 

 \begin{figure}%[h]  
    \centering
\includegraphics[width=\linewidth]{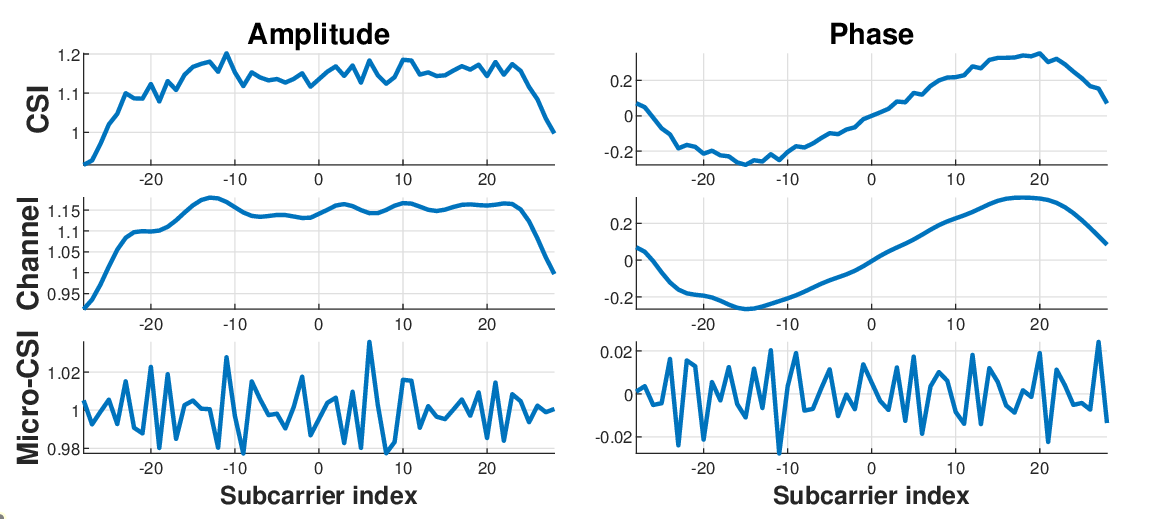}
  \caption{Micro-CSI extracted from denoised LoS CSI measurements\cite{kong2023physicallayer,kong2024csirff}.}
  \label{fingerprint}
\end{figure}

Previous studies have established that it is viable to identify wireless devices based on characteristic imperfections in their RF circuitry, such as sampling frequency offset, transient signals, I/Q imbalance, nonlinearity of power amplifier, etc. \cite{ transient, danev2009physical, Brik,huang2016specific,nguyen2011device,hao2014relay,vo2016fingerprinting}. Given the challenges in explicitly and accurately modeling the subtle nonlinear distortions intrinsic to RF hardware, deep learning (DL) emerges as an intuitive solution for RFF identification and authentication systems. Recent advances have proven that DL, thanks to their multidimensional mappings, can achieve fingerprinting accuracy levels impossible to achieve by traditional low-dimensional algorithms \cite{riyaz2018deep,gopalakrishnan2019robust,hanna2020deep,soltani2020more,cekic2021wireless,xie2021generalizable,youssef2018machine,hanna2019deep,meneghello2022deepcsi,yang2023eliminating,das2018deep,roy2019rfal,yu2019robust,restuccia2019deepradioid,sankhe2019oracle,al2020exposing}.  
The introduction of deep learning has boosted the ability to accurately identify devices with the presence of environmental changes \cite{restuccia2019deepradioid,al2020exposing}. However, the abovementioned fingerprinting methods require access to physical-layer I/Q samples and additional instruments for I/Q sample collection, which largely limits their practical application in COTS systems.

Recent advancements in WiFi channel state information (CSI) tools \cite{atheros,linux} have enabled COTS WiFi devices to convey physical-layer CSI measurements to higher layers, which, in turn, has enhanced the feasibility of CSI-based RF fingerprinting (CSI-RFF) \cite{hua,liu,lin2020,kong2023physicallayer,kong2024csirff,meneghello2022deepcsi,yang2023eliminating}. Compared to approaches reliant on physical layer I/Q samples, CSI-RFF offers a more efficient and streamlined deployable alternative. A number of studies \cite{hua,liu,lin2020,kong2023physicallayer,kong2024csirff} have engaged model-based techniques to distill various CSI-derived fingerprints, yielding commendable levels of accuracy. Nevertheless, these methodologies generally require a sizable number of CSI measurements to mitigate noise and channel distortions for reliable identification or authentication. Furthermore, these models are typically constrained by their training and testing environments, either being evaluated within identical environmental conditions \cite{hua,liu,lin2020} or within new environments under only line-of-sight (LoS) scenarios \cite{kong2023physicallayer,kong2024csirff}. Recent approaches \cite{meneghello2022deepcsi,yang2023eliminating} utilize deep learning to extract RF fingerprints from CSI. However, authors in \cite{meneghello2022deepcsi} report a significant degradation in performance when exposed to unseen channel conditions. While the work \cite{yang2023eliminating} has shown resilience against environmental dynamics, such as human movement, the stationary positioning of devices within their experiments raises questions about the generalizability of their method under more diverse and fluctuating channel conditions. Hence, the dynamic nature of wireless channels and the presence of noise on subtle fingerprints represent substantial challenges to the robustness of CSI-RFF systems.
 
In this work, we propose DeepCRF, a deep learning-enhanced CSI-RFF framework designed for channel-resilient and robust WiFi device identification, leveraging micro-signals on CSI curves, termed micro-CSI, as distinctive RF fingerprints. Our prior research \cite{kong2023physicallayer,kong2024csirff} has established the efficacy of micro-CSI in RF fingerprinting and developed a signal space-based (SS) method that effectively extracts micro-CSI for RFF in strong LoS conditions. 
However, our recent experiments, detailed in Sec.~\ref{why}, have revealed that the SS method struggles to cope with the complexities of non-line-of-sight (NLoS) conditions, failing to extract micro-CSI effectively. This limitation has motivated the design of DeepCRF, which utilizes a carefully trained convolutional neural network (CNN) to extract robust features (or representations) of micro-CSI across diverse channel conditions. The primary contributions of this study are three-fold:
\begin{itemize}[leftmargin=*]
    \item %We systematically evaluate the impact of wireless channels on the accurate extraction of micro-CSI to facilitate the design of DeepCRF. 
    A primary challenge in designing DeepCRF is to ensure that it learns small-scale micro-CSI (i.e., the RF fingerprint), not large-scale channel information, given the variability of wireless channels. To tackle this, we introduce a model-inspired data augmentation technique tailored for micro-CSI, enabling the network to learn it while reducing extensive data collection requirements. Furthermore, to counter the noise impact on the small-scale micro-CSI, we employ supervised contrastive learning to enhance the noise robustness of the learned representations. 
    \item We carefully design the neural network structure of DeepCRF for complex-valued CSI input data. The design of DeepCRF is the result of an extensive exploration of architectural choices. This exploration led to a structure that not only outperforms state-of-the-art neural network structures \cite{meneghello2022deepcsi,yang2023eliminating} but does so with a more efficient parameterization.  Besides, we address the challenge of noise on small-scale micro-CSI by examining various fusion methods, including data fusion and 3 types of decision fusion methods. The fusion methods are integrated within the DeepCRF framework, enhancing its robustness against noise, with an average of 10\% performance improvement in real-world evaluation. 
    \item  We undertook a comprehensive evaluation of DeepCRF, using a controllable synthetic testing dataset and an extensive real-world dataset\footnote{The collected real-world dataset and the code are publicly available at \url{https://github.com/Oriseven/DeepCRF_TIFS}.}. The latter was compiled by gathering CSI from 19 COTS WiFi 4/5/6 network interface cards (NICs) positioned across various positions. Our results from both synthetic and real-world datasets indicate that DeepCRF demonstrates superior performance in indoor environments, covering both LoS and NLoS conditions, compared to the signal space-based method and other neural network-based baselines. Notably, it achieves an average identification accuracy of 99.53\% in real-world unseen scenarios using 4 CSI measurements per identification procedure. 
\end{itemize}
{Compared to the conference version \cite{kong2024channelresilient}, this work refines the neural network architecture of DeepCRF to achieve improved accuracy and parameter efficiency in device identification. Moreover, we develop and compare two data augmentation methods and integrate a decision fusion technique to enhance the system's robustness against noise. We also evaluate the performance of DeepCRF using a real-world indoor dataset, where it outperforms state-of-the-art CSI-based RFF methods.}

\section{Related Work}
\subsection{Model-based RFF} Initial endeavors in RFF concentrated on the extraction of unique fingerprints from the transient components of transmitted signals \cite{ transient}. However, the practical application of transient-based RF fingerprints is constrained due to their sensitivity to the positioning of the device and the orientation of the antenna's polarization \cite{transient}. The attention is then directed toward the extraction of steady-state features in pursuit of more stable characteristics from the received signals. This includes features such as power spectral density \cite{danev2009physical}, synchronization correlation \cite{Brik}, non-linearities of the power amplifier (PA) \cite{huang2016specific}, carrier frequency offset (CFO) \cite{nguyen2011device}, I/Q imbalances \cite{hao2014relay}, and sampling frequency offsets \cite{vo2016fingerprinting}. While promising, these methods necessitate the acquisition of physical-layer I/Q samples, often requiring specialized instrumentation for collecting I/Q samples. 
A recent trajectory of research has investigated the extraction of RF fingerprints from CSI, utilizing model-based methods \cite{hua, liu, lin2020, kong2023physicallayer,kong2024csirff}. For instance, the extraction of CFO from CSI measurements is proposed in \cite{hua}. However, CFO as a fingerprint might not be viable in scenarios involving mobility. Further developments include the work of \cite{liu}, which presented methods for deriving nonlinear phase errors from CSI, and \cite{lin2020}, which focused on the extraction of power variances induced by PA from CSI. Additionally, our prior work \cite{kong2023physicallayer,kong2024csirff} proposed the use of micro-signals present in CSI as unique identifiers. Nonetheless, there remains a significant gap in the literature, as these approaches have yet to enhance the robustness of fingerprinting systems to channel dynamics.

\subsection{DL-based RFF} Deep learning has emerged as a promising tool in RFF systems, achieving higher accuracy in device identification than model-based approaches. Researchers have explored a wide array of data representations, including raw I/Q samples \cite{riyaz2018deep,gopalakrishnan2019robust,sankhe2019oracle,hanna2020deep,soltani2020more,cekic2021wireless,xie2021generalizable}, Fourier and Wavelet transforms  \cite{youssef2018machine,hanna2019deep}, and CSI measurements \cite{meneghello2022deepcsi,yang2023eliminating}, in conjunction with diverse neural network architectures such as multilayer perceptrons \cite{youssef2018machine}, CNN \cite{riyaz2018deep,hanna2019deep,youssef2018machine,hanna2020deep, xie2021generalizable}, and recurrent neural network \cite{das2018deep, roy2019rfal} to test their effectiveness across various channel conditions. Notable studies have demonstrated substantial accuracies in device fingerprinting, yet the challenge of channel variation impacts remains \cite{yu2019robust,restuccia2019deepradioid,al2020exposing,soltani2020more,xie2024radio}. Some work has suggested introducing synthetic impairments to improve model robustness, though the literature still lacks a clear explanation on how they can be compensated at the receiver \cite{sankhe2019oracle, riyaz2018deep}. Further innovation has been seen with the use of complex-valued CNNs, which align well with the complex nature of I/Q samples, indicating a promising direction for refining deep learning methods in RFF \cite{gopalakrishnan2019robust, cekic2021wireless}.

The studies most closely aligned with our work are \cite{meneghello2022deepcsi, yang2023eliminating}, which explore the utilization of CSI to diminish the requisite hardware complexity for sample acquisition. The authors in \cite{meneghello2022deepcsi} have applied CNNs with integrated attention mechanisms to distill radio frequency fingerprints from compressed beamforming feedback. Their approach achieves a 98\% identification accuracy across 10 WiFi radios. Nonetheless, the accuracy of their method shows a marked decrease under conditions of channel variations, with the accuracy diminishing to 47.28\%. 
Later, the authors in \cite{yang2023eliminating} have identified and addressed the issue of fingerprint fracture, a phenomenon indicative of the vulnerability inherent in model-based feature extraction \cite{liu}. Their proposed solution leverages an autoencoder network to effectively address environmental interference affecting the phase error drift range. The refined fingerprint is subsequently fed into a CNN endowed with attention mechanisms to identify access points with a 100\% success rate by using 100 CSI measurements for each identification procedure. However, their evaluation was limited as the devices were positioned at a fixed location, which does not account for the potential challenges posed by location changes and device mobility.

\section{Preliminaries, Motivation, and Problem Formulation}\label{sec:pre}
In this section, we first present the signal model of micro-CSI formulated in our prior work \cite{kong2023physicallayer,kong2024csirff}. We then provide justification, through preliminary experiments, for the necessity of deep learning in achieving channel-resilient micro-CSI-based RFF. Finally, we describe the RFF identification problem within the context of deep learning. 

\subsection{Signal Model of Micro-CSI}
The estimated CSI at the receiver side incorporates the distortions induced by both wireless channels and the transmitter’s RF circuitry imperfections\footnote{{Here, we assume that the receiver, as the CSI collector, will not be replaced. As such, the RF circuitry imperfection induced by the receiver can be treated as a constant.}}, with the latter often being used as RF fingerprints. As shown in our prior work \cite{kong2023physicallayer,kong2024csirff}, hardware distortions can be modeled as deviations $\bold{d}$ from standard channel training symbols $\bold{t}$. Under this model, the time-domain signal emitted to the air can then be written as $\bold{s=t+d}$. The received signal can be written as $\bold{y=h \ast s + z}$, where $\ast$ denotes circular convolution, $\bold{h}$ is the discrete-time equivalent channel, and $\bold{z}$ is complex white Gaussian noise. The frequency-domain CSI $\tilde{\bold{c}}$ estimated by least squares (LS) algorithm \cite{book80211} can be expressed as:  
\begin{equation}\label{est_CSI}
\tilde{\bold{c}}=\tilde{\bold{y}}\circ\tilde{\bold{t}}=\tilde{\bold{h}}\circ(\bold{1}+\tilde{\bold{d}}\circ\tilde{\bold{t}})+\tilde{\bold{z}} = \Tilde{\bold{h}} \circ (\bold{1}+\Tilde{\bold{f}})+\Tilde{\bold{z}},
\end{equation}
where $\tilde{(\cdot)}$ represents signal in the frequency domain, $\Tilde{\bold{f}}=\tilde{\bold{d}}\circ\Tilde{\bold{t}}$ denotes the hardware imperfection-induced fingerprint, $\circ$ is element-wise multiplication and $\bold{1}$ represents the vector with all 1's. We remark that prevailing RFF frameworks often model hardware distortions as a function $f(\bold{s})$ applying to the transmitted symbols $\bold{s}$ \cite{hanna2020deep, xie2021generalizable}. In contrast, our approach exploits the properties of CSI estimation, which utilizes the Long Training Symbols (LTS), a predefined sequence of symbols in the preamble of each transmitted WiFi packet. This protocol-defined consistency suggests that hardware distortions would produce uniform deviations to the standard LTS transmitted by a specific device. As a result, we are allowed to directly estimate the fingerprint vector $\Tilde{\bold{f}}$ from CSI measurements of the said device, circumventing the need to fully characterize the distortion function $f(\cdot)$.

As shown by Eq. (\ref{est_CSI}), the channel information $\Tilde{\bold{h}}$ and the hardware fingerprint $\Tilde{\bold{f}}$ are entangled together, making the extraction of $\Tilde{\bold{f}}$ nontrivial. As an initial attempt, our prior work \cite{kong2023physicallayer} employs the signal space-based (SS) method to estimate the fingerprint ${\Tilde{\bold{f}}}$, termed micro-CSI due to its small scale, from denoised CSI under strong LoS conditions, as depicted in Fig.~\ref{fingerprint}. The denoising process involves averaging multiple CSI measurements gathered in a static environment over a short period. 
The key insight behind the framework developed in \cite{kong2023physicallayer,kong2024csirff} is that under strong LoS, the wireless channels are inherently sparse, and thus the channel information $\Tilde{\bold{h}}$ only occupies a non-significant number of dimensions (taps) after converting the CSI measurement $\Tilde{\bold{c}}$ into time domain through inverse Fourier transform. In this case, the fingerprint $\Tilde{\bold{f}}$ residing in other unoccupied dimensions can be separated from the channel information. A very recent study \cite{yang2024led} introduces a discrete wavelet transform (DWT) method to extract RF fingerprints from the demodulation reference signal (DMRS) in Long-Term Evolution (LTE) systems. This approach is based on the assumption that multipath interference predominantly affects the low-frequency components of the CSI. Recognizing its potential for micro-CSI extraction, we have incorporated this method into our subsequent analysis.

\subsection{Why Deep Learning?}
\label{why}

\begin{figure*}
\subfloat[SS \cite{kong2023physicallayer,kong2024csirff}]{\includegraphics[width=0.5\linewidth]{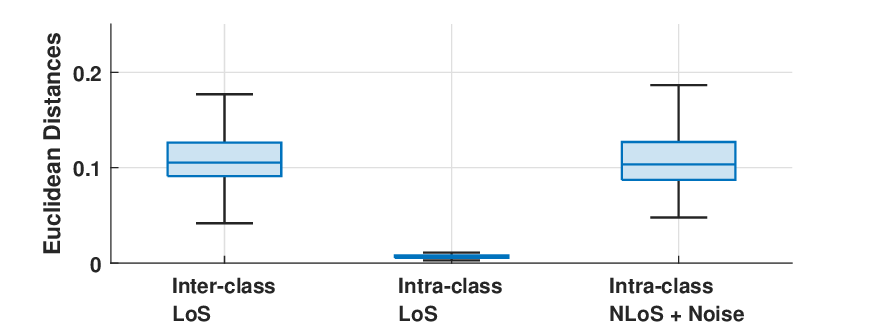}}
\subfloat[DWT \cite{yang2024led}]{\includegraphics[width=0.5\linewidth]{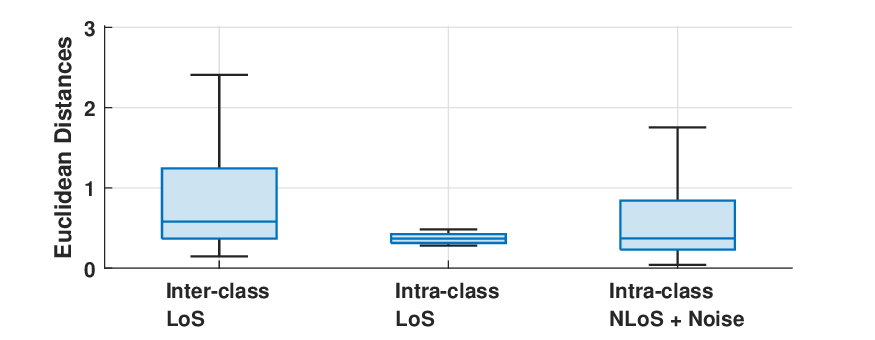}}
   \caption{Distances between fingerprints of different NICs and fingerprint extraction errors caused by environmental changes.}
   \label{distance}
\end{figure*}

In this subsection, we conduct a preliminary study to explore the potential limitations of the SS \cite{kong2023physicallayer,kong2024csirff} and DWT \cite{yang2024led} methods using real-world CSI measurements collected under complex multipath and mobile conditions. To that end, we first apply the SS and DWT methods to extract fingerprints from CSI measurements. Subsequently, we calculate the pairwise Euclidean distances between the extracted fingerprints. Fig.~\ref{distance} presents our findings using box plots to illustrate the inter-class and intra-class distances among the extracted fingerprints. The inter-class distances refer to the distances between fingerprints from different network interface cards (NICs), highlighting the identifiability between NICs. On the other hand, intra-class distances, which are derived from CSI pertaining to a single NIC subject to various channel conditions, characterize the stability of fingerprints despite channel condition changes.

\begin{table}%[h]
  \caption{WiFi Devices}
  \label{devices}
  \centering
  \resizebox{0.85\linewidth}{!}{
  \begin{tabular}{ccccc}
    \toprule
    NIC No.&WiFi&Brand&Model&Quantity\\
    \midrule
    C1-C6 &WiFi 4 & MediaTek  & MT7601 & 6\\
    C7 &WiFi 4 &MediaTek& MT7612& 1\\
    C8 &WiFi 4 & Realtek& RT3070& 1\\
    C9-C15 &WiFi 4& Atheros & AR9271& 7\\
    C16 &WiFi 5 & Realtek& RT8811 & 1\\
    C17 &WiFi 5 & Realtek& RT8821 & 1\\
    C18 &WiFi 5 & Realtek& RT8822 & 1\\
    C19 &WiFi 6 & Intel& AX200 & 1\\
  \bottomrule
\end{tabular}
}
% \vspace{-1em}
\end{table}

In our study, we tested 19 WiFi NICs involving NICs from the same and different models, as listed in Table~\ref{devices}. These 19 NICs serve as transmitters to be identified, supported by the same host PC in turn at the same location under strong LoS conditions to the receiver. For each NIC, we used 100 consecutive CSI measurements from one receiving antenna for each fingerprint construction. In Fig.~\ref{distance}, we present box plots of inter-class and intra-class under LoS conditions. The first two box plots in Fig.~\ref{distance}a show that intra-class variations under LoS conditions are less than the minimum distinguishable distance between 19 distinct WiFi NICs. This observation confirms the effectiveness of the SS method in LoS environments. Note that the observed variability in intra-class distances under LoS conditions can be primarily attributed to residual noise, which cannot be completely eliminated during the denoising process. However, in Fig.~\ref{distance}b, the intra-class distances under LoS conditions slightly exceed the minimum distinguishable distance. This means the DWT method might not be able to completely eliminate channel effects, which could pose challenges in differentiating devices, even under LoS conditions. The superior performance of the SS method stems from its ability to more precisely eliminate channel effects by leveraging the properties of the pulse shaping filter.

We then explored the impact of multipath and noise on intra-class distances. Specifically, we used the CSI measurements collected in an NLoS scenario with an average signal-to-noise ratio (SNR) of 28 dB. The third box plot in both Fig.~\ref{distance}a\&b shows the variance of intra-class distance between fingerprints extracted from CSI measurements collected in the presence of multipath and noisy channel conditions. Specifically, to visualize the detrimental impact of noise on the extraction process, each fingerprint was extracted from a single CSI measurement without the denoising process. The fingerprints extracted under NLoS and noisy channels exhibit fluctuations that exceed the minimum distinguishable distance for reliable differentiation among NICs. This observation indicates that for the SS and DWT methods, the variability induced by the channel conditions could result in the misidentification of a single NIC as multiple distinct entities. It also underscores the adverse impact that multipath channels have on the accuracy of CSI-RFF using the SS and DWT methods. An intuitive explanation of this observation is that the presence of more multipath makes more taps in the time domain or higher frequency in the frequency domain being used by the channel. Therefore, it is harder for the SS and the DWT methods to extract accurate fingerprints. Further, the small scale of micro-CSI makes it particularly susceptible to noise.

As a quick summary, multipath and noise pose some challenges to the model-based method for micro-CSI extraction. Besides, any inaccurate pre-processing may cause a loss of RF information in CSI. These challenges motivate our design of DeepCRF, a model-inspired deep learning approach for micro-CSI-based fingerprinting by using raw CSI measurements. 

\subsection{Identification Problem in DeepCRF}
The goal of identification is to determine the identity of a wireless device from a pool of potential candidates, utilizing the collected CSI $\Tilde{\bold{c}}$. Within the context of deep learning, this identification process typically involves the use of a feature (or representation) vector $\bold{r}$, representative of the RFF learned from the collected CSI, rather than using the CSI itself. 
The RFF identification can be formulated as a classification problem given by
\begin{equation}
\begin{gathered}
\hat{m} = \underset{m \in \mathbf{M}}{\text{argmax}} \left( \hat{y}_{m} \right) \\
\text { subject to } \quad \hat{\mathbf{y}}=C\left(\mathbf{r}\right),
 \mathbf{r}=E\left(\Tilde{\mathbf{c}}\right),
\end{gathered}
\end{equation}
where $\hat{m}$ is the predicted identity (or class). The set of all potential classes is denoted by $\mathbf{M}$, while $\hat{y}_m$ represents the predicted likelihood that the input CSI $\Tilde{\mathbf{c}}$ is associated with class $m$, and $\hat{\mathbf{y}}$ is the resultant vector of probabilities across all classes. 
$E(\cdot)$ is the extraction function used to learn the RF representations from the CSI, referred to as the RFF extractor. $C(\cdot)$ is the classification function used to classify devices from the learned representations, referred to as the RFF classifier.

The efficacy of the extractor $E(\cdot)$ and classifier $C(\cdot)$ is contingent upon the quality of the training process. Denote the training dataset as $\mathscr{T} = \{(\Tilde{\bold{c}}^{(n)}, \bold{y}^{(n)})\}_{n=1}^{N}$, which encompasses $N$ pairs of training samples. 
Here, $\Tilde{\bold{c}}^{(n)}$ represents the $n$-th CSI instance, and $\bold{y}^{(n)}$ denotes the corresponding true identity, encoded as a one-hot vector. The objective of the training is to minimize the expected loss across the training dataset, formulated as the following optimization problem
\begin{equation}
\min \mathbb E_{\mathscr{T}} ~\mathcal{L}\left(\hat{\mathbf{y}}^{(n)},\mathbf{y}^{(n)}\right) 
\end{equation}
where $\mathcal{L}(\cdot)$ represents the loss function used to evaluate the difference between the predicted identity vector $\hat{\mathbf{y}}^{(n)}$ and the ground-truth identity vector $\mathbf{y}^{(n)}$.

\section{DeepCRF Design}
\label{method}
This section first overviews the neural network structure of DeepCRF before elaborating on the design of each component and the training strategies.

\subsection{Overview of DeepCRF}

{The extractor $E(\cdot)$ and the classifier $C(\cdot)$ within the DeepCRF framework are trained using a two-stage process, as illustrated in Fig.~\ref{overview}. Given the variability of wireless channels across different environments, it is critical to ensure that $E(\cdot)$ learns the intrinsic RF distortions present in the CSI rather than those distortions induced by the wireless channel. To achieve this, we have developed a model-inspired data augmentation strategy specifically for the RF fingerprint under consideration, namely micro-CSI. This augmentation process introduces additional variability and diversity into the training dataset, thereby enhancing the generalization capability of DeepCRF to handle unseen channel conditions.} 

During Training Stage 1, we adopt a supervised contrastive loss function \cite{khosla2020supervised} to enhance noise resistance. This loss function prompts the extractor to generate RF representations that are closer for CSI inputs of the same class, while maintaining further distance between RF representations from different classes. The inclusion of a project head, corroborated by previous research to be beneficial \cite{chen2020simple}, provides auxiliary guidance throughout the training process. In Training Stage 2, our goal is to train a classifier to realize robust device identification. The pre-trained parameters of the extractor are transferred, providing a foundational understanding for further refinement. The focus then shifts to train the classifier. {Concurrently, the extractor is fine-tuned to synergize with the classifier.} Throughout Stage 2, the cross-entropy loss function is used to guide the training process. This two-stage approach ensures that DeepCRF not only learns representative features against noise but also achieves high accuracy in device identification.

\subsection{{Model-Inspired Data Augmentation}}
\label{dataaugmentation}

\begin{figure}
    \centering
\includegraphics[width=\linewidth]{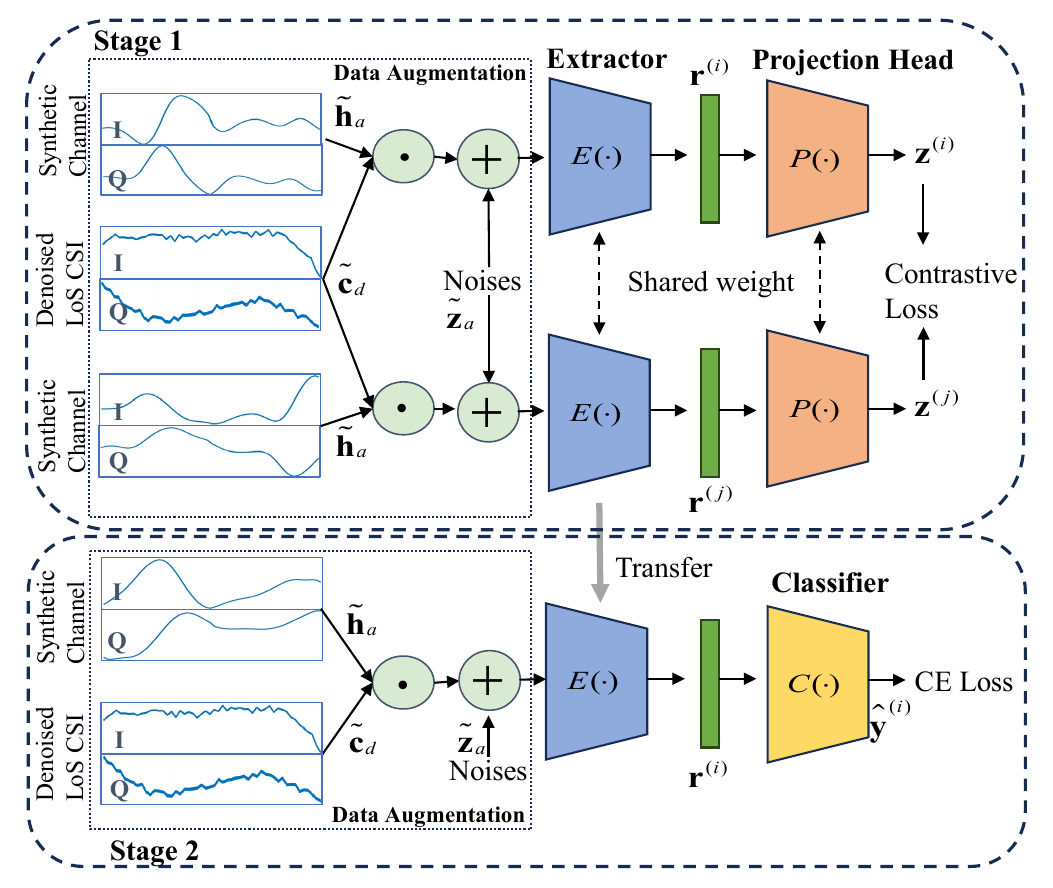}
    \caption{Overview of DeepCRF.}
    \label{overview}
% \vspace{-1em}
\end{figure}

In order for the extractor to effectively learn the representations of RF fingerprints, it is critical to expose it to a sizeable dataset spanning diverse channel conditions. However, gathering a large-scale practical CSI dataset can be a time-consuming process. To circumvent this problem, we devise an augmentation strategy based on the signal model of micro-CSI. This approach allows us to synthesize augmented CSI samples from a relatively small amount of real-world CSI measurements collected from a few positions in a room. For clarity, henceforth, we will distinguish real-world CSI measurements from their synthetic counterparts by referring to the latter as \textit{CSI samples}. 

Let $\Tilde{\bold{c}}_d$ denote the denoised CSI measurements. Assuming that noise is negligible in the denoised CSI, we can express the denoised CSI as $\Tilde{\bold{c}}_d \approx \Tilde{\bold{h}} \circ (\bold{1}+\Tilde{\bold{f}})$. In this expression, it is the hardware-induced distortions (i.e., $\Tilde{\bold{f}}$) from which we desire the extractor to learn representative features. By contrast, the channel-induced distortion (i.e., $\Tilde{\bold{h}}$) fluctuates with environmental changes and lacks the consistency needed to serve as a reliable identifier across different devices. Therefore, the augmentation strategy should artificially introduce a wide range of channel conditions while preserving the integrity of $\Tilde{\bold{f}}$. This approach enhances the extractor's ability to learn device-specific RF features, minimizing the impact of channel-induced variations. Furthermore, as discussed in Section~\ref{why}, the sensitivity of micro-CSI to noise, due to its small-scale nature, underscores the necessity to include noise in the augmentation process.

Given the above considerations, we could have two data augmentation strategies. The first strategy involves component-wise multiplication of the fingerprints $(\bold{1}+\hat{\Tilde{\bold{f}}})$ estimated by the SS method \cite{kong2023physicallayer,kong2024csirff} under LoS scenarios with synthetic channels $\Tilde{\bold{h}}_a$ generated using simulation software such as MATLAB. The reliability of the SS method under LoS conditions ensures that the extracted fingerprint preserves the integrity of $\Tilde{\bold{f}}$. The synthetic CSI samples can be expressed~as
\begin{equation}
\label{aug0}
\Tilde{\bold{c}}_{a,1}= \Tilde{\bold{h}}_a \circ (\bold{1}+\hat{\Tilde{\bold{f}}}) + \Tilde{\bold{z}}_a \approx \Tilde{\bold{h}}_a \circ (\bold{1}+\Tilde{\bold{f}}) + \Tilde{\bold{z}}_a,
\end{equation}
where $\Tilde{\bold{z}}_a$ is the added synthetic noise. 

By contrast, the second strategy is to component-wise multiply the denoised CSI $\Tilde{\bold{c}}_d$ with synthetic channels, which can be formulated as 
\begin{equation}
\label{aug}
\Tilde{\bold{c}}_{a,2}= \Tilde{\bold{h}}_a \circ \Tilde{\bold{c}}_d + \Tilde{\bold{z}}_a \approx (\Tilde{\bold{h}}_a \circ \Tilde{\bold{h}}) \circ (\bold{1}+\Tilde{\bold{f}}) + \Tilde{\bold{z}}_a.
\end{equation} 
It is evident that the synthesized samples $\Tilde{\bold{c}}_a$ in both methods align with the structure of the signal model defined in \eqref{est_CSI} and maintain the integrity of the fingerprint. 
It is important to note that in the real-world, the discrete-time equivalent channel $\Tilde{\bold{h}}$ captures the combined effects of multipath propagation and filtering \cite{taubock2010compressive}. This channel can be described as $\Tilde{\bold{h}} = \Tilde{\bold{h}}_c \circ \Tilde{\bold{h}}_p$, where $\Tilde{\bold{h}}_c$ denotes the physical channel and $\Tilde{\bold{h}}_p$ represents the filtering effect. The 802.11 standard specifies a spectral mask to guide the implementation of pulse-shaping filtering, which requires attenuation of the signal to certain levels at designated frequency offsets from the center frequency \cite{9442429}. This spectral mask imposes characteristics distinct from those of the physical channel itself.  
Therefore, in this work, we opted for the second strategy that uses denoised CSI $\Tilde{\bold{c}}_d$ for data augmentation as $\Tilde{\bold{c}}_d$ contains not just the fingerprint but also the filtering effects that are inherently present in real-world CSI measurements, thereby enhancing the ability to generalize to real-world environments. By contrast, in the first strategy, the filtering effects have been removed during the fingerprint extraction process. Fig.~\ref{overview} offers a visual representation of the model-inspired data augmentation process. Later, in Section \ref{impact}, we will evaluate and demonstrate the enhanced identification accuracy achieved by employing the second augmentation strategy compared to the first strategy.

\begin{figure}

\subfloat[Architecture]{\includegraphics[width=\linewidth]{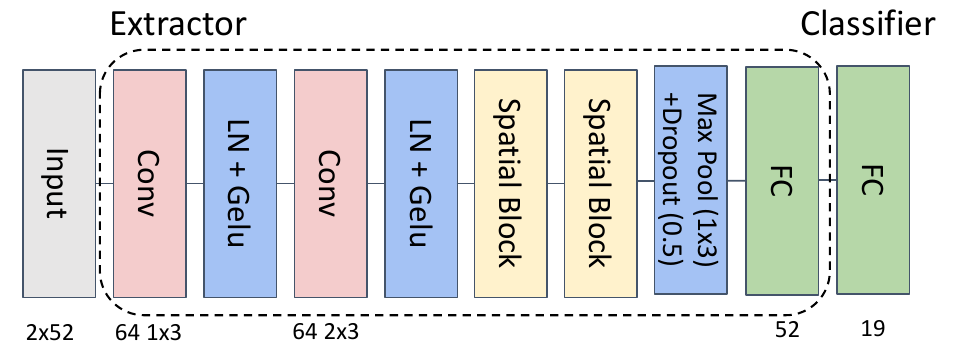}}
\hfil
\subfloat[Spatial Block]{\includegraphics[width=\linewidth]{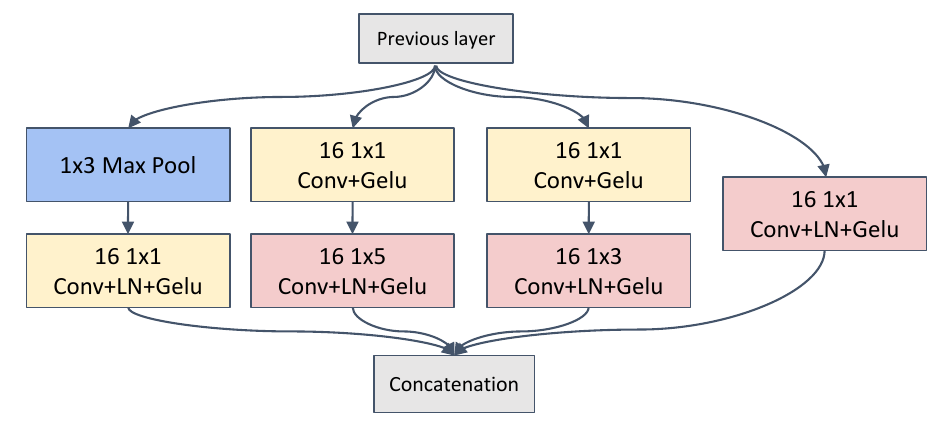}}
   \caption{Building Blocks of DeepCRF.}
   \label{building_block}
\end{figure}

\subsection{Neural Network Design}
 \label{netowrk_design}
We now delve into the designs of the three neural network components of DeepCRF. 

The extractor $E(\cdot)$ is responsible for transforming the CSI input, either $\Tilde{\bold{c}}$ or $\Tilde{\bold{c}}_a$, into a latent representation $\bold{r}$. That is,  $\bold{r}= E(\Tilde{\bold{c}}) \text{ or } E(\Tilde{\bold{c}}_a)$. Considering the similarity of CSI to image data rather than time-series data, we adopt a convolutional neural network architecture as the foundation for our extractor. While the ideal method for processing complex-valued CSI data would involve complex-valued neural networks, these networks are associated with high model complexity and significant increases in model size \cite{gopalakrishnan2019robust, cekic2021wireless}. To balance performance and complexity, we have carefully designed a real-valued neural network for the extractor, aiming to achieve performance comparable to complex-valued networks but with reduced complexity. The efficiency of our designed real-valued extractor will be evaluated and demonstrated in Section \ref{impact}. Before feeding the complex-valued CSI input into the real-valued extractor, it is decomposed into two real-valued vectors. One vector represents the real (I) part of the CSI input, and the other captures the imaginary (Q) part of the CSI input. This decomposition process is illustrated in Fig.~\ref{overview}.

The proposed extractor consists of two convolutional layers, two spatial blocks, and a fully connected (FC) layer, as depicted in Fig. \ref{building_block}(a). Each convolutional layer is characterized by three parameters: the first parameter represents the number of filters, and the second and third parameters specify the size of the filter. In the first convolutional layer, 64 filters of size $1\times3$ are employed, allowing each filter to learn spatial information in the frequency domain over either the real or imaginary dimension independently. Recognizing the interrelationship between the real and imaginary parts of the wireless channel in CSI, the second convolutional layer employs 64 filters, each of size $2\times3$. The second layer is designed to capture information across both real and imaginary dimensions from the feature maps generated by the first convolutional layer, enabling the network to exploit the inherent dependencies.

Capturing both local and global features in CSI is important, as RF distortions are distributed across all subcarriers. Drawing inspiration from the inception network \cite{inception}, spatial blocks are designed at the third and fourth layers of the proposed network to enable it to learn spatial patterns at different scales. Each spatial block comprises multiple convolutions with varying filter sizes, namely $1\times1$, $1\times3$, and $1\times5$, allowing the network to capture diverse spatial patterns across different scales. 
Besides, our experimental observations have revealed that employing the Gaussian Error Linear Unit (GELU) as the activation function, using Layer Normalization (LN), and reducing the number of normalization layers contribute to improved network performance. These design choices and their benefits also find support in \cite{liu2022convnet}. Fig.~\ref{building_block}(b) provides a detailed illustration of the spatial block's architecture. Furthermore, the last FC layer serves the purpose of feature reduction. The number of neurons in this layer is determined by the number of subcarriers, which is 52 in our case. This design choice ensures that the extracted features are compact and representative.

Another component of DeepCRF is the projection head $P(\cdot)$ within Training Stage 1. It translates the learned representations $\bold{r}$ into a hypersphere where the contrastive loss is applied for better alignment and uniformity of representations \cite{wang2020understanding}. We use a dense layer featuring 52 neurons to obtain the normalized output as $\bold{z} = \frac{P(\bold{r})}{||P(\bold{r})||_2}$ \cite{chen2020simple}. In the second stage, the classifier $C(\cdot)$, implemented as a single dense layer, is responsible for mapping the representations $\bold{r}$ to their corresponding class probabilities $\hat{\mathbf{y}}$, expressed as $\hat{\mathbf{y}}= C(\bold{r})$. The number of neurons is determined by the total number of classes present in the identification task, which is 19 in our experiments. 

\subsection{Loss Functions}
 
We now specify the design of loss functions in DeepCRF. In Training Stage 1, our objective is to train an extractor that can learn a reliable RF representation $\mathbf{r}$ from CSI input, rather than trying to precisely quantify the hardware imperfections, $\Tilde{\bold{f}}$. This approach is driven by the overlapping nature of channel effects, $\Tilde{\bold{h}}$, and hardware imperfections, $\Tilde{\bold{f}}$, in the time domain, making an exact estimation of $\Tilde{\bold{f}}$ impractical. Consequently, our methodology is designed to extract a feature vector, $\mathbf{r}$, that effectively represents $\Tilde{\bold{f}}$ and captures the crucial information needed for accurate device identification. To guide the training process in Stage 1, we employ supervised contrastive loss \cite{khosla2020supervised}. This loss function encourages the extractor to learn discriminative representations that effectively distinguish between different transmitters while maintaining similarity among representations of the same transmitter. 
The contrastive loss $\mathcal{L}_{\text{con}}$ is computed as the sum of individual contrastive losses $\mathcal{L}^{(i)}_{\text{con}}$ for each arbitrary CSI input $i$ in the batch $I$. The contrastive loss for a specific CSI input $i$ is defined as: 
\begin{equation}
\label{loss_eq}
\mathcal{L}^{(i)}_{\text{con}} = \frac{-1}{|J(i)|} \sum_{j \in J(i)} \log \frac{\exp({\bold{z}^{(i)}}^T \bold{z}^{(j)} / \tau)}{\sum_{k \in K(i)} \exp({\bold{z}^{(i)}}^T \bold{z}^{(k)} / \tau)},
\end{equation}
where $K(i) \equiv I \backslash  \{i\}$, the term $J(i)$ represents the set of indices of all samples in $K(i)$ with the same label as $i$, and $|J(i)|$ is the cardinality of $J(i)$. 
In Training Stage 2, the model is guided by cross-entropy (CE) loss, which is a widely used objective function for multi-class classification problems. The CE loss $\mathcal{L}_{\text{ce}}$ is computed as the mean of individual CE losses, $\mathcal{L}^{(i)}_{\text{ce}}$, for each arbitrary CSI input $i$ in the batch $I$. Formally, the CE loss for a specific CSI input $i$ can be defined as:
\begin{equation}
\mathcal{L}^{(i)}_{\text{ce}} = - \sum_{m \in \mathbf{M}} y^{(i)}_m \log(\hat{y}^{(i)}_m)  %(\mathbf{y}_i, \hat{\mathbf{y}}_i)
\end{equation}
where $\mathbf{M}$ denotes the set of all possible classes, $y^{(i)}_m$ represents the true probability of the $i$-th CSI input belonging to class $m$, and $\hat{y}^{(i)}_m$ denotes the predicted probability of the $i$-th CSI input belonging to class $m$.

\subsection{Noise Compensation}
\label{fusion_method}
As highlighted in Section~\ref{why}, the small-scale nature of micro-CSI makes it highly susceptible to noise, necessitating the implementation of robust noise mitigation strategies to enhance the system performance. To suppress the noise effect in DeepCRF, we employ the concept of collective decision-making by leveraging multiple CSI inputs from the same device. This approach can be realized through either data fusion or decision fusion techniques \cite{Combination}. Data fusion involves the integration of multiple raw CSI data to denoise the information before making a decision, while decision fusion entails combining the decisions made from individual CSI inputs. 
In our study, we opted for the decision fusion approach because decisions processed through the neural network exhibit greater consistency than raw data, resulting in enhanced fusion performance. To methodically assess the effectiveness of various decision fusion techniques, we experimented with multiple fusion methods, including Majority Voting (MV), Average Probabilities (AP), and Borda Count (BC) \cite{Combination}. After rigorous performance evaluations, as detailed in Section~\ref{impact}, we adopted the AP fusion method due to its superior performance across our tests.

The AP method fuses decisions by averaging the predicted probabilities from multiple CSI inputs of the same device and selecting the class with the highest average probability as the final decision. Mathematically, the AP fusion can be expressed~as 
\begin{equation}
    \hat{m}_{\text{avg}}= \underset{m \in \mathbf{M}}{\text{argmax}} \left( \sum_{n=1}^{N_{c}} \hat{y}^{(n)}_{m} \right)
\end{equation}
where $\hat{m}_{\text{avg}}$ represents the final decision based on the average probabilities, and $N_{c}$ is the total number of CSI instances used for fusion. 
The experimental findings in \cite{kong2024csirff} show that differences between micro-CSIs extracted from different receiving chains of the same receiver are negligible. This consistency across receiving chains facilitates the use of CSI instances from multiple receiving antennas to enhance the fusion process. Consequently, $N_{c} = N_{csi} \times N_{rx}$, where $N_{csi}$ represents the number of CSI measurements, and each measurement comprises $N_{rx}$ CSI instances obtained from $N_{rx}$ receiving antennas. 
By averaging the probabilities, the AP fusion method effectively mitigates the impact of noise and helps to smooth out any outliers or erroneous predictions, resulting in a more stable and accurate final decision. 

\section{Data Collection and Augmentation}
In this section, we begin by detailing the CSI collection setup and the realization of our data augmentation strategy. Subsequently, we delve into the specifics of datasets splitting and training implementation.

\subsection{Real-world Data Collection}
  \begin{figure*}
  \centering
  \subfloat[Room A, Jul 2023]{\includegraphics[width=0.31\linewidth]{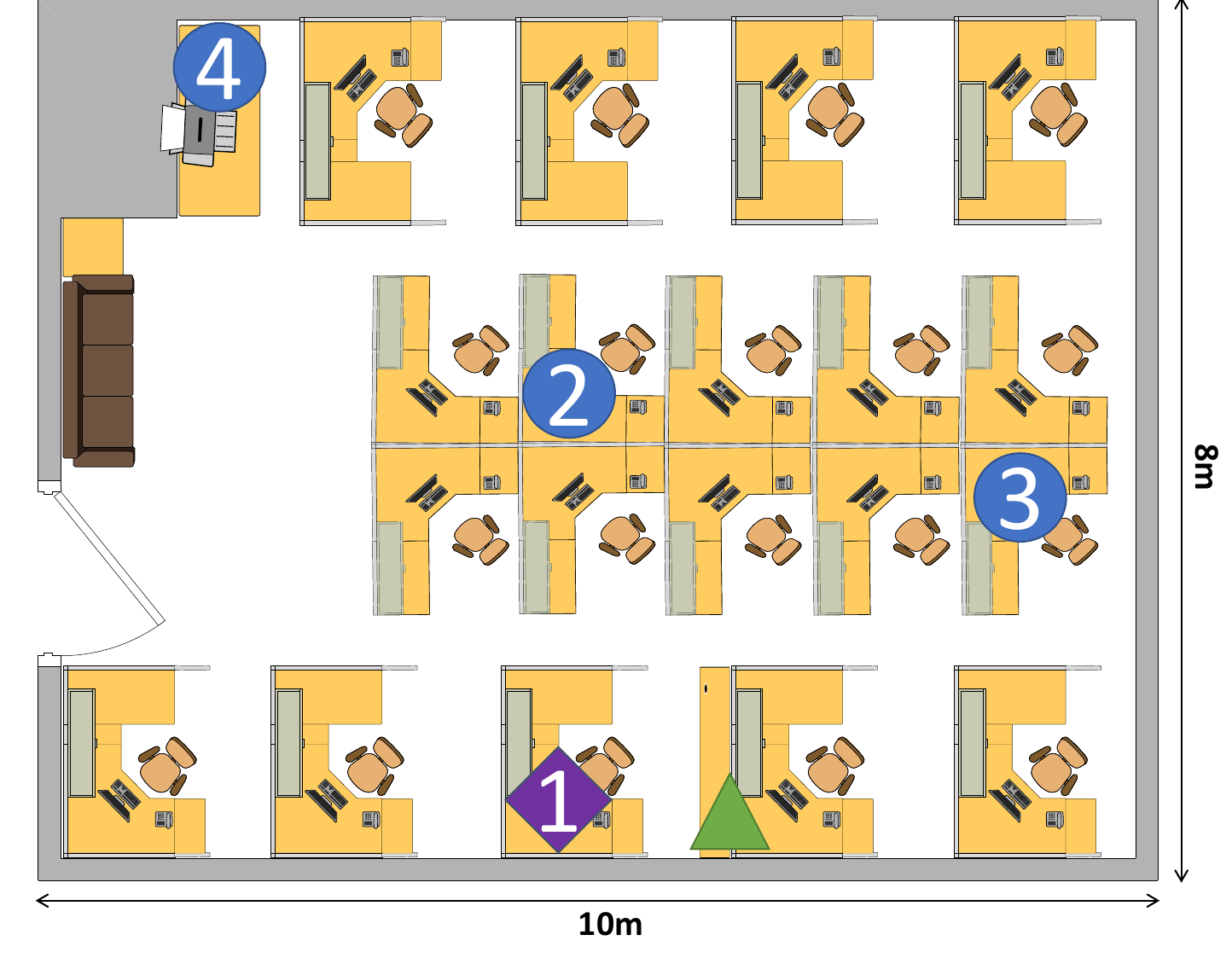}}
\subfloat[Room B, Nov 2023]{\includegraphics[width=0.20\linewidth]{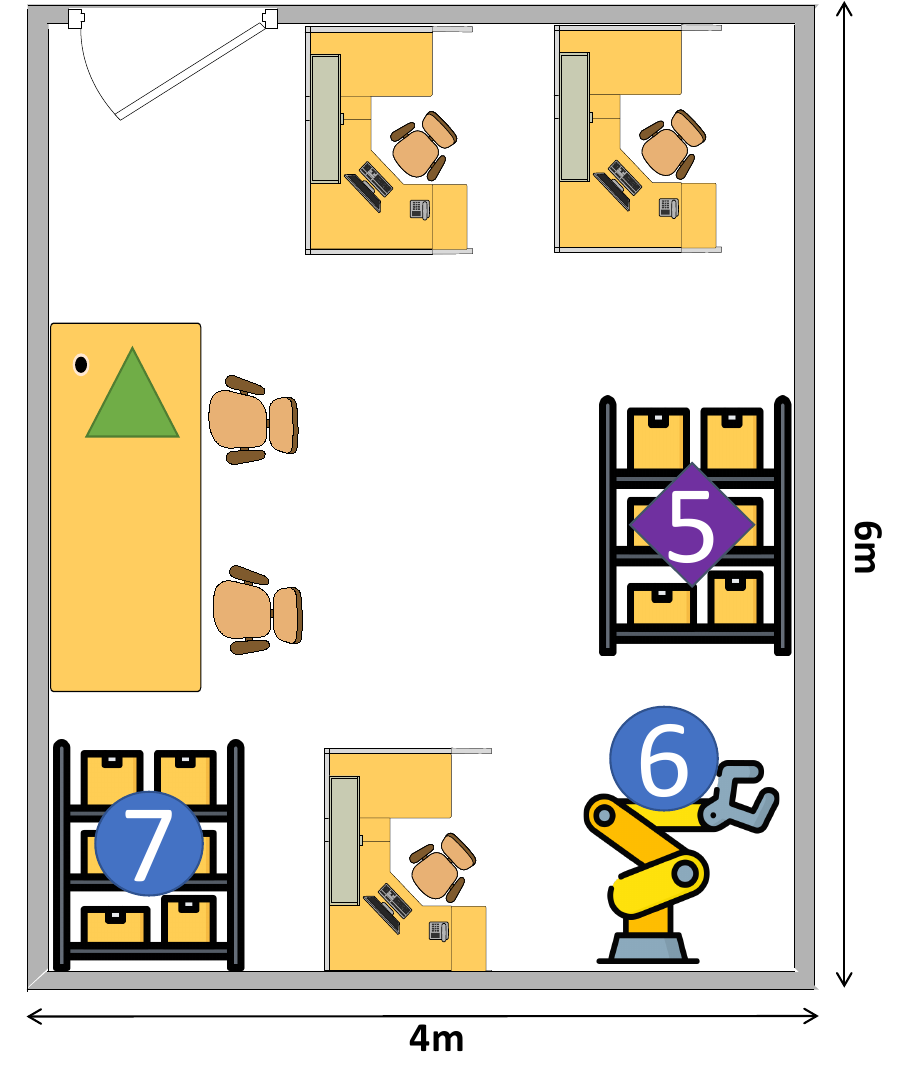}}
\subfloat[Outdoor, Jul 2024]{\includegraphics[width=0.20\linewidth]{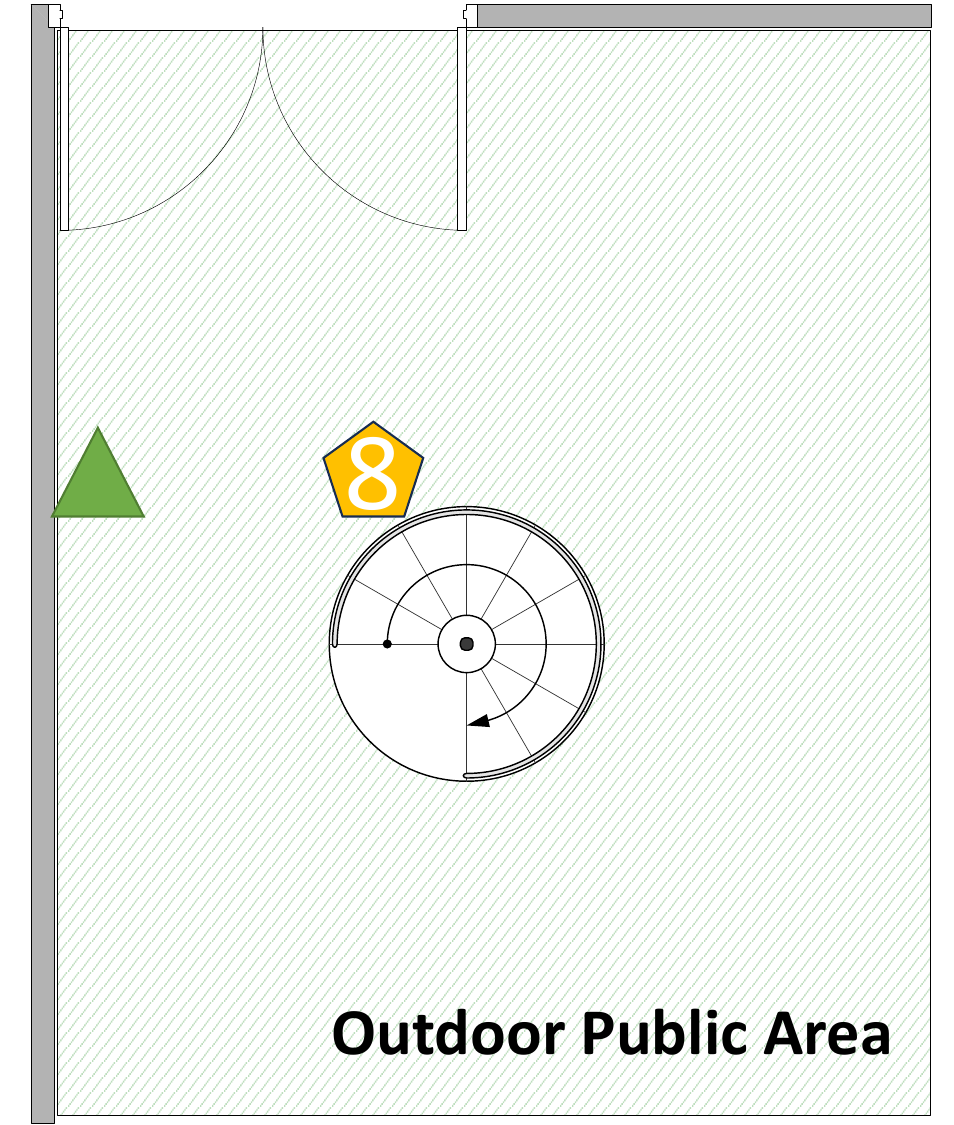}}
\subfloat[Car Park, Jul 2024]{\includegraphics[width=0.28\linewidth]{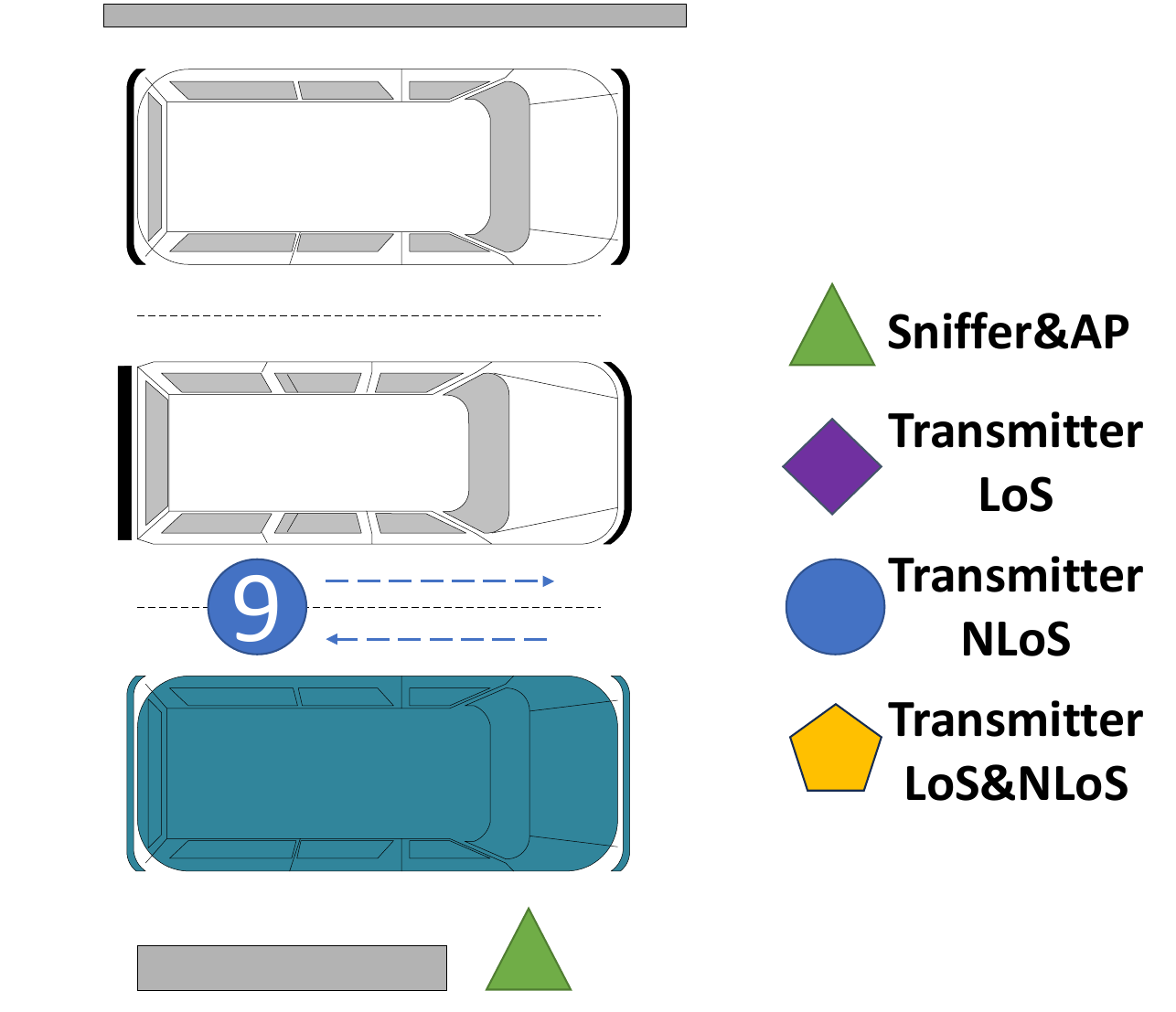}}
\caption{Experiment environments and details.}
\label{setup}
     % \vspace{-1em}
   \end{figure*}

\begin{figure}
    \centering
\includegraphics[width=\linewidth]{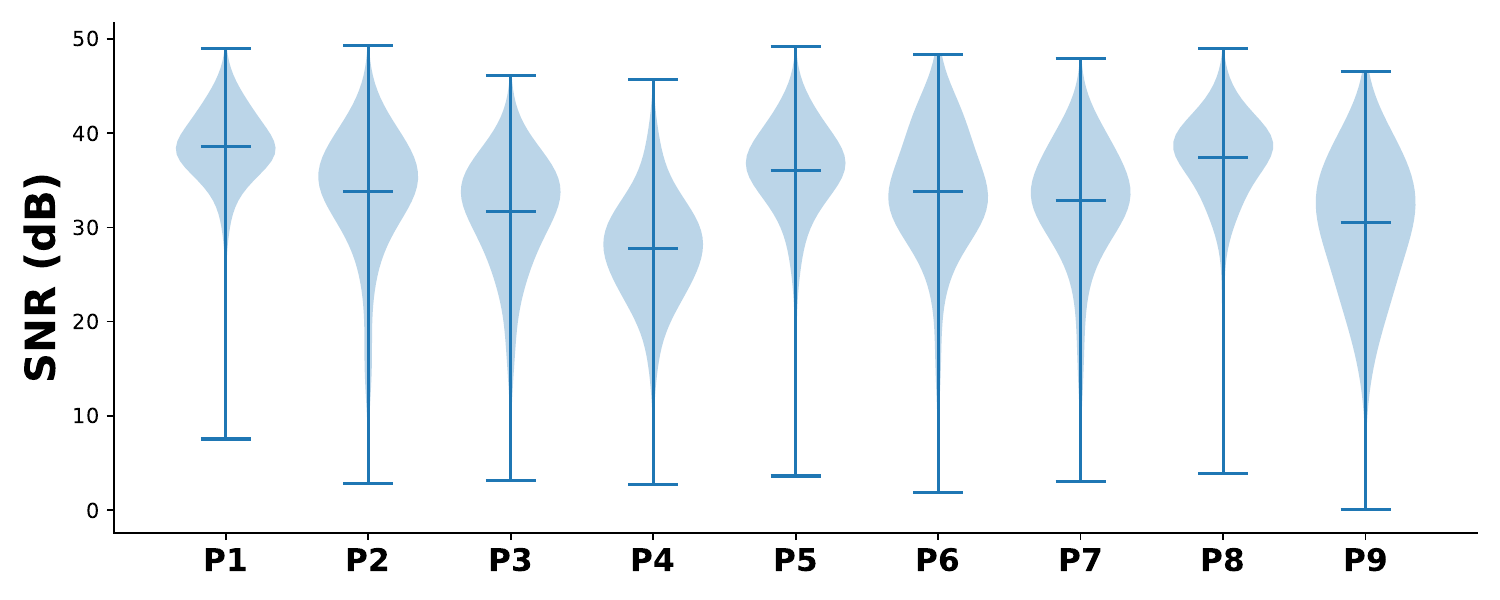}
    \caption{SNR distribution of collected data in different positions.}
    \label{snr}
% \vspace{-1em}
\end{figure}
After a thorough investigation, we have discovered that existing public CSI datasets exhibit several limitations that render them unsuitable for our training and testing needs. These limitations include small dataset sizes and a restricted number of positions from which the CSI measurements are collected. To address these limitations and ensure the robustness and effectiveness of DeepCRF, we have collected our own CSI dataset specifically for the subsequent performance evaluation experiments. Our dataset comprises CSI measurements from 19 WiFi 4/5/6 devices sourced from 4 different manufacturers, as detailed in Table \ref{devices}. Specifically, our device set includes the most challenging scenario: identifying devices within the same model, which may exhibit similar manufacturing distortions. 

\textbf{CSI Collection Toolkit}: Our previous work \cite{kong2023physicallayer} adopted a popular CSI Toolkit \cite{atheros} compatible with commodity WiFi devices for CSI collection, which provided a validation of the micro-CSI from popular CSI Toolkits. However, the used CSI Toolkit is only available for Atheros NICs, which restricts its generalizability across different brands. In this work, we have chosen to employ CSI derived from ACK (acknowledgment) packets for several reasons. First, our experimental findings indicate that ACK packets consistently use the legacy format (802.11a/g), which includes 52 subcarriers regardless of the WiFi protocol version. This consistency in CSI length ensures uniform input data, facilitating the implementation of our model across different versions of WiFi protocols. Second, RFF identification frameworks typically require multiple CSI measurements for a single identification. Given their small data size and short transmission duration, ACK packets present a significant advantage by minimizing the consumption of transmission resources. 
Despite the advantages, current open-source toolkits do not support CSI extraction from ACK packets. Consequently, we have adopted a software-defined radio (SDR) approach for our evaluations\footnote{A recent study \cite{esp} confirmed the feasibility of extracting CSI from ACK packets using the commodity ESP32 platform. However, the toolkit used in \cite{esp} has not yet been made open-source.}. Specifically, the used SDR consists of a Xilinx Zynq ZC706 development \cite{zc706} and an FMCOMMS5 ADI daughter board with four antennas \cite{coms5}, $N_{rx} = 4$. The acquired samples were decoded, and their corresponding CSI reports were extracted using a modified MATLAB 802.11 analysis program \cite{analysis}. 

\textbf{Network Configuration}: The network was randomly configured to work on the 2.4GHz channel 10 with a bandwidth of 20 MHz (i.e., the center frequency is 2457 MHz) and run the 802.11n protocol. Every NIC listed in Table \ref{devices} was installed on the same mini-PC to respond with ACK in turn, where only one device under identification was transmitting at any time. Installed on the same mini-PC assured that the fingerprint difference is solely caused by the NICs. Besides, the laboratory settings were typical network-heavy environments, potentially including interference from other devices.

\textbf{Data Collection Scenarios}: In our study, data collection was thoughtfully conducted in two distinct indoor environments, an outdoor area, and a car park scenario, to ensure a comprehensive analysis of DeepCRF's performance across different environments. The first room, Room A, is a $10m \times 8m$ research office, which is complex with many obstacles around. The clapboards around each desk construct a complex multipath environment. 
The second room, Room B, is a $6m \times 5m$ common room and presents a relatively simpler multipath channel. The common room accommodates a few desks, a cabinet, and a robot arm. Data was gathered from seven specific locations within these rooms during busy office hours, representing a medium level of human activity. These locations were deliberately chosen to include both LoS and NLoS scenarios, as depicted in Fig.~\ref{setup}a and Fig.~\ref{setup}b. The third scenario is an outdoor public area in front of the main entrance of a building on our campus, as depicted in Fig.~\ref{setup}c. People were frequently moving in and out, often walking around or passing directly between the transceiver devices. This scenario provided us with both LoS and NLoS data amidst significant human activity, with the NLoS data resulting from human body blockages. Last, we gathered data in a car park where the transmitter was mobile, carried by an individual moving around a car, with the transceiver positioned in an NLoS situation relative to the transmitter, as depicted in Fig.~\ref{setup}d. The presence of the car as a substantial obstacle and the occasional movement of other vehicles around the setup further enriched our dataset with challenging signal obstruction scenarios. The numbers on the icons in Fig.~\ref{setup} represent the position indices in our dataset. 
To provide a detailed insight into the signal quality at each location, the SNR information was visualized using violin plots in Fig.~\ref{snr}. These plots conveyed the distribution of SNR values for each position and also highlighted the corresponding maximum, minimum, and mean values of SNR. 

\textbf{Real-World Dataset}: The data collection process involved conducting 30 rounds of CSI measuring for each NIC in each position. To ensure temporal diversity, a time gap of 30 seconds was maintained between each two collections in each position. During each CSI collection round, the NICs transmitted ACK packets to the access point. These ACK packets use 52 active subcarriers. However, due to the nature of wireless communication, packet losses occurred, resulting in slightly different amounts of CSI measurements acquired by the CSI collector for each NIC. 
To thoroughly assess the robustness of DeepCRF, we collected data in four environments in separate months (i.e., July 2023, November 2023, and July 2024). This approach allows us to capture the spatial and temporal variations in the CSI measurements. The resulting real-world dataset comprises a total of 879,943 CSI measurements, and we partition the real-world dataset into three distinct subsets:
\begin{itemize}[leftmargin=*]
    \item Training set: 80\% of the CSI measurements from positions P1 (LoS) and P2 (NLoS) are allocated for training\footnote{The decision to train the classifier with data from both a LoS scenario and an NLoS scenario was strategically made to encompass a comprehensive range of micro-CSI intensity variations, as discussed in Section VII-A.}. It is important to note that the training set of the real-world dataset will be combined with the synthetic training set, which will be detailed in Section~\ref{syntheticdata}. 
    \item Validation set: 10\% of the CSI measurements from positions P1 and P2 are designated for validation. Similar to the training set, the validation set of the real-world dataset will also be combined with the synthetic validation set during the training process. 
    \item Testing set: The remaining 10\% of the CSI measurements from positions P1 and P2, along with all CSI measurements from positions P3 to P9, are reserved for testing. This subset allows us to evaluate the performance of DeepCRF on unseen data and assess its ability to generalize to different positions and environments.
\end{itemize}

\subsection{Synthetic Dataset for Data Augmentation}
\label{syntheticdata}

Collecting a large-scale dataset of practical CSI measurements under diverse environments and from various NICs is a laborious task. To overcome this challenge and enhance the training process, we use a small amount of real-world CSI measurements to synthesize augmented CSI samples. Specifically, we use 80\% of the CSI measurements from two positions, P1 and P2 in Fig.~\ref{setup}a, which represent LoS and NLoS scenarios, respectively, to generate the synthetic dataset. 

\begin{table}%[h]
  \caption{WLAN Indoor Channel Models \cite{Erceg2004IEEEPW}}
  \label{channel}
  \resizebox{\linewidth}{!}{
  \begin{tabular}{ccccc}
    \toprule
    Channel&RMS delay&Number of &Number of& \multirow{2}{*}{Environment}\\
    Model& spread (ns)&clusters&taps\\
    \midrule
    B &15   & 2 & 9 &Residential\\
    C &30 & 2& 14&Small Office\\
    D &50 & 3& 18&Typical/Large 
Office\\
  \bottomrule
\end{tabular}
}
% \vspace{-1em}
\end{table}

As a preprocessing step, we first denoise the CSI measurements from P1 and P2 using the techniques discussed in Section~\ref{why}. To expand the diversity of the real-world dataset, we augment the denoised CSI measurements by simulating six different types of indoor channel conditions. These conditions include both LoS and NLoS scenarios for three indoor channel models: Model-B, Model-C, and Model-D. We use the \texttt{wlanTGnChannel} function from the WLAN toolbox \cite{wlan} to simulate these channel conditions. The delay spreads and cluster parameters of these channel models are listed in Table~\ref{channel}. The resulting synthetic dataset comprises 2,188,800 CSI samples, which encompass 45,600 realizations of the six specified channel types. To further diversify the dataset, we introduce noise across eight SNR levels, ranging from 5 dB to 40 dB, using the \texttt{awgn} function. %This step helps to incorporate the impact of varying noise levels on the CSI measurements.

For our experimental evaluations, we partition the synthetic dataset into three distinct subsets: 80\% of the samples are allocated for training, 10\% for validation, and the remaining 10\% for testing. 
By leveraging this synthetic dataset, we can train our model on a diverse range of channel conditions and noise levels, enhancing its ability to handle realistic scenarios.

\subsection{Training Details}
We employ the Adam optimizer, which is a popular choice for training deep learning models due to its adaptive learning rate and efficient convergence properties. The initial learning rate is set to $10^{-3}$, and we apply a weight decay of $10^{-4}$ as a regularization technique. The batch size is set to 512, according to our empirical evidence. Furthermore, we use a patience value of 10 to tolerate non-converging validation loss. For the loss function defined in \eqref{loss_eq}, we set the temperature parameter $\tau$ to 0.07, following the recommendation from \cite{khosla2020supervised}. The proposed architecture is implemented in PyTorch and executed on an Ubuntu platform using an NVIDIA RTX A6000 GPU equipped with 48 GB of VRAM. Each training epoch of the model was completed in approximately 24 seconds.

\section{Evaluation}

In this section, we conduct a comprehensive evaluation of the identification performance of DeepCRF using both synthetic and real-world datasets. Moreover, we investigate the impact of different design choices and training strategies. Additionally, we visualize the learned features to gain insights into the discriminative power of DeepCRF. 

\subsection{Baselines}
The baseline methods selected for comparison encompass two latest neural network architectures tailored for CSI-based RFF identification, i.e., Self-ACC \cite{yang2023eliminating} and DeepCSI \cite{meneghello2022deepcsi}, and the signal space-based (SS) method developed in our prior work \cite{kong2023physicallayer,kong2024csirff} for fingerprint extraction. To ensure a fair comparison, we implemented the SS method with an identical classifier architecture to that of DeepCRF and trained it using the same real-world dataset for device identification. Furthermore, we preserved the architectural designs of Self-ACC and DeepCSI but enhanced them with our proposed data augmentation (DA) strategy used in DeepCRF to evaluate the efficacy of the neural networks under enhanced conditions. 
The adjustment made to DeepCSI involved using raw CSI as the input instead of the compressed beamforming feedback (CBF) used in \cite{meneghello2022deepcsi}. This is because our dataset, which is collected in a single transmit antenna setup, does not support the computation of CBF. We remark that raw CSI offers more comprehensive information than CBF and thus may potentially enhance the performance of DeepCSI.

\begin{figure*}
    \centering
    \subfloat[Channel B LoS]{\includegraphics[width=0.32\linewidth]{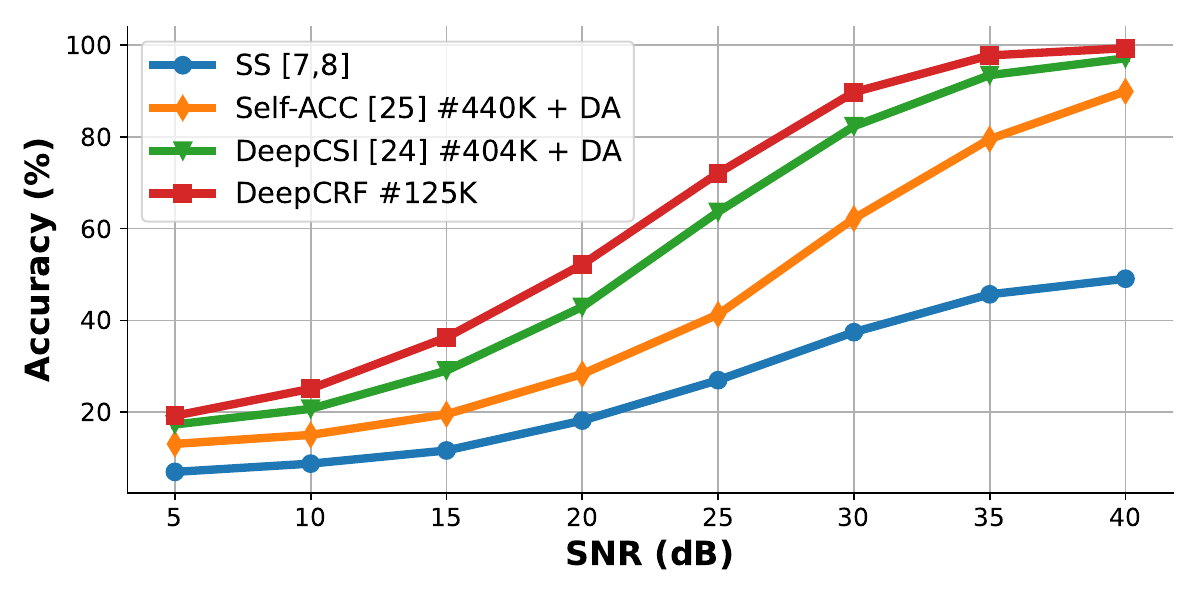}}
    \subfloat[Channel C LoS]{\includegraphics[width=0.32\linewidth]{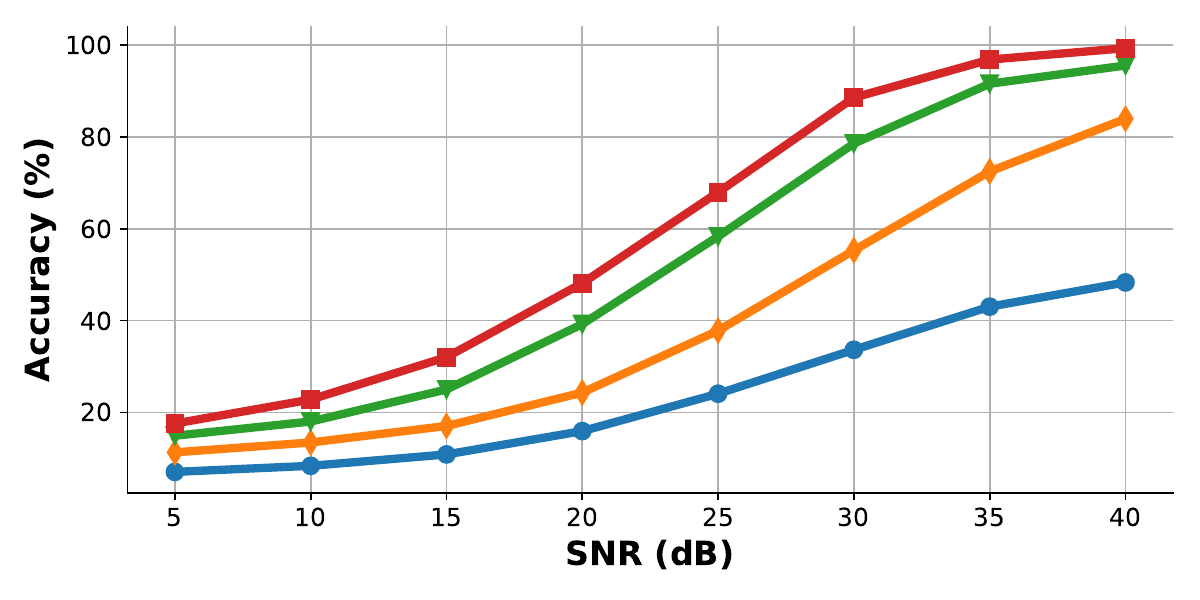}}
    \subfloat[Channel D LoS]{\includegraphics[width=0.32\linewidth]{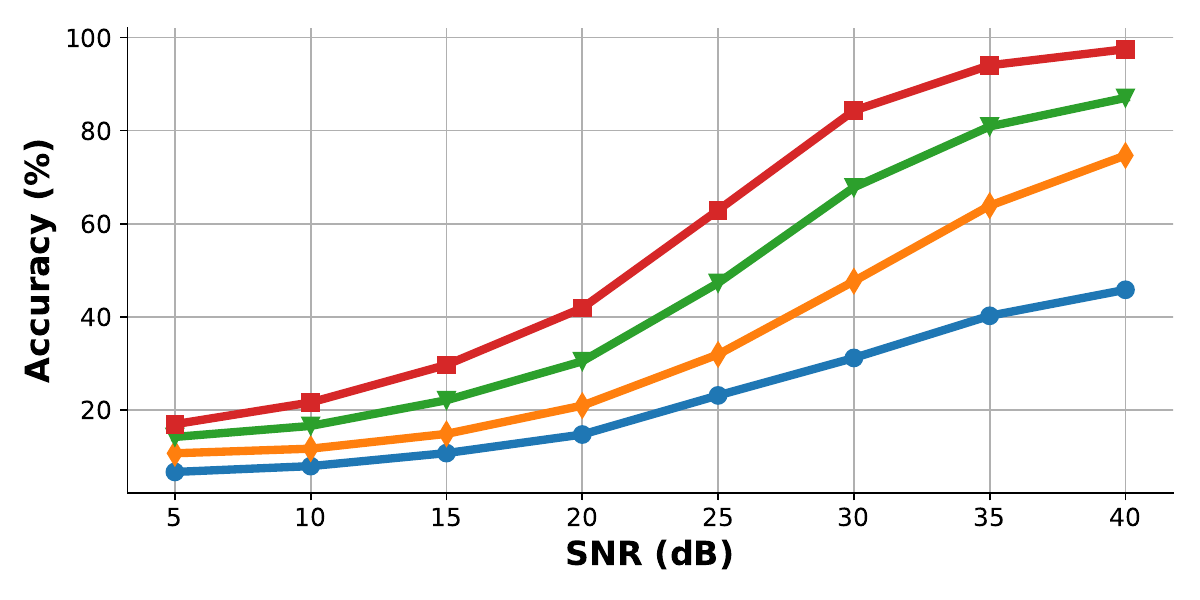}}
    % \vspace{-1em}
    \hfil
    \subfloat[Channel B NLoS]{\includegraphics[width=0.32\linewidth]{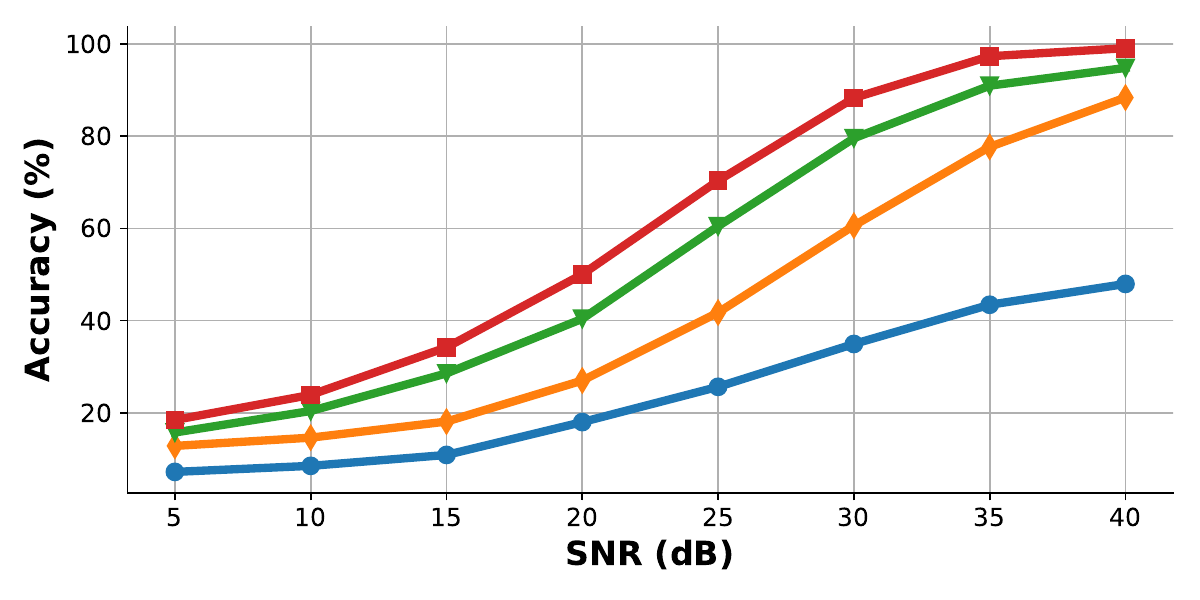}}
    \subfloat[Channel C NLoS]{\includegraphics[width=0.32\linewidth]{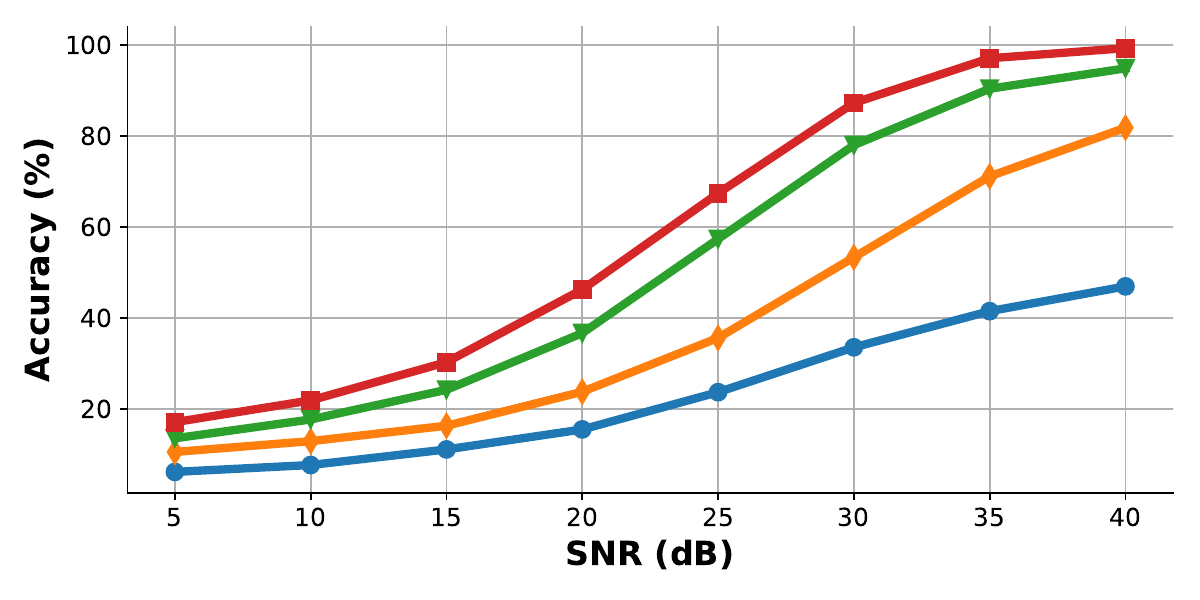}}
    \subfloat[Channel D NLoS]{\includegraphics[width=0.32\linewidth]{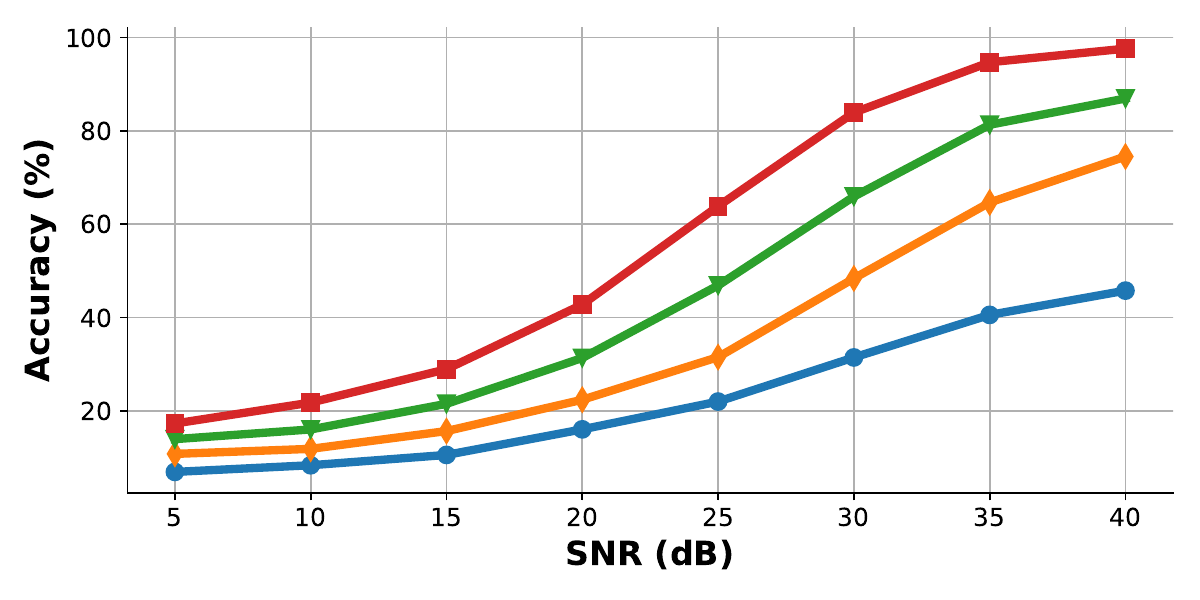}}
    
\caption{Performance comparison of DeepCRF with other neural network architectures (Self-ACC \cite{yang2023eliminating}, DeepCSI \cite{meneghello2022deepcsi}) and the SS method \cite{kong2023physicallayer,kong2024csirff} under different channel models and SNR values using synthetic CSI data, when $N_{c} = 1$. In the legend, DA represents data augmentation.}
\label{model}
% \vspace{-1em}
\end{figure*}

\subsection{{Identification Performance}}
We now assess the identification performance of DeepCRF using both synthetic and real-world datasets, comparing it with the selected baselines.

\textbf{Performance on Synthetic Dataset}: We begin by comparing the performance of DeepCRF and other baselines on the synthetic dataset using a single CSI sample for each identification. The synthetic dataset includes CSI samples under a range of challenging and previously unseen channel conditions at various SNR levels. The performance of DeepCRF with other baseline methods under different channel models and SNR values is presented in Fig.~\ref{model}. The average accuracy among 19 NICs is reported for each method across varying SNR levels ranging from 5 dB to 40 dB. The results highlight the superior performance of DeepCRF compared to the other methods, particularly the SS method. The SS method exhibits subpar performance across all tested SNR levels, even in high SNR conditions. For instance, at an SNR of 40 dB, the SS method achieves an accuracy of only around 47\% within all channel types. This degraded performance suggests that the SS method is highly vulnerable to multipath interference, which impairs its ability to accurately extract fingerprints. 

Among the deep learning-based methods, DeepCRF stands out with its superior performance. In all channel model scenarios, DeepCRF surpasses both DeepCSI and Self-ACC, achieving an average accuracy increase of $7.7\%$ and $19.5\%$, respectively. 
Further, as the complexity of the channel increases from channel model B to C and D, the performance gap between DeepCRF and the two baselines widens. This observation highlights the superior ability of DeepCRF to handle more complex multipath scenarios.  
In the NLoS scenarios shown in Figs.~\ref{model}d-\ref{model}f, DeepCRF still maintains its superior performance compared to the other methods. We also observe that there is a slight decrease in accuracy compared to the LoS cases, which is expected due to the increased complexity in NLoS environments. 
Furthermore, it is worth noting that DeepCRF achieves its superior performance while using the least number of parameters compared to Self-ACC and DeepCSI. Specifically, DeepCRF has only 125 thousand parameters, while Self-ACC and DeepCSI have 440 and 404 thousand parameters, respectively. This parameter efficiency of DeepCRF highlights its ability to learn highly discriminative features with a more compact network architecture.

As observed in Fig.~\ref{model}, the performance deterioration in NLoS scenarios is marginal for each channel model (B, C, and D), with an approximate decrease of only 1\% compared to their respective LoS scenarios. The minimal performance gap between LoS and NLoS conditions under the same channel model can primarily be attributed to the similar number of paths and path range maintained across both conditions within each channel model. This consistency suggests that the presence or absence of a dominant direct path does not significantly impact performance when the overall path characteristics remain similar. Instead, what markedly influences performance is the degree of multipath complexity.

\begin{figure*}
    \centering
    \subfloat[$N_{csi} = 1, N_{rx} = 1$]{\includegraphics[width=\linewidth]{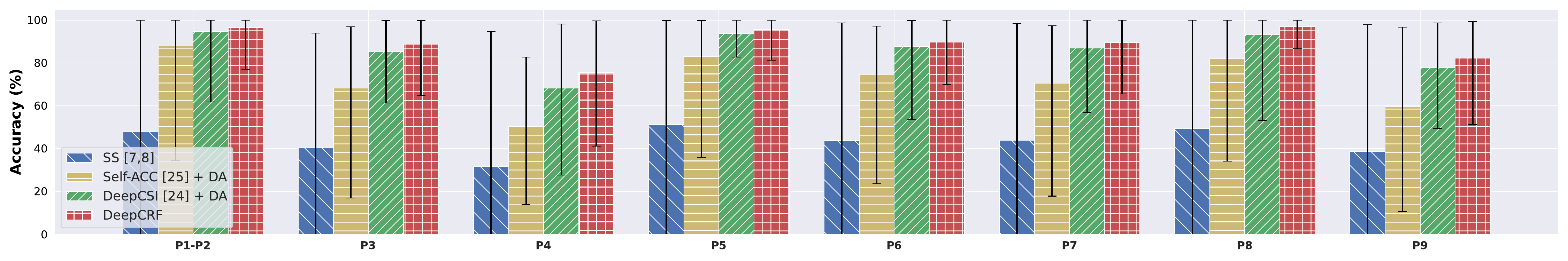}}
    % \vspace{-1em}
    \hfill
    \subfloat[$N_{csi} = 4, N_{rx} = 4$]{\includegraphics[width=\linewidth]{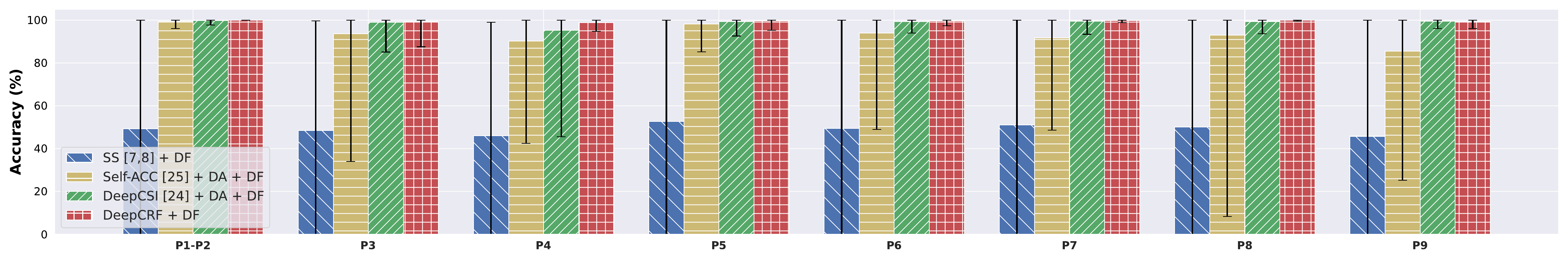}}
    \caption{Performance comparison of DeepCRF with other neural network architectures (Self-ACC \cite{yang2023eliminating}, DeepCSI \cite{meneghello2022deepcsi}) and the SS method \cite{kong2023physicallayer,kong2024csirff} under different positions using real-world CSI measurements.In the legend, DA and DF represent data augmentation and decision fusion, respectively.}
    \label{models_practical}
% \vspace{-1em}
\end{figure*}

\textbf{Performance on Real-World Dataset}: 
Next, we shift the evaluation to the real-world dataset. Fig.~\ref{models_practical} presents the performance comparison of DeepCRF with the other methods under different positions. The accuracy of each method is evaluated at nine positions (i.e., P1 to P9). 
The bar plot in Fig.~\ref{models_practical}a shows the average accuracy of each method across the 19 NICs at each position, using a single CSI measurement for each identification. Additionally, error bars are included to indicate the highest and lowest accuracy achieved among the 19 NICs. To illustrate the impact of decision fusion (DF) on the performance of each method, Fig.~\ref{models_practical}b includes accuracy values for each method when the number of CSI measurements used for DF ($N_{csi}$) is equal to 4. By comparing the accuracy performance in Fig.~\ref{models_practical}a and Fig.~\ref{models_practical}b, we can observe the gain of the DF method.

From Fig.~\ref{models_practical}a, we can observe that DeepCRF maintains its superior performance in real-world scenarios, achieving the highest accuracy across all positions. Self-ACC and DeepCSI exhibit lower accuracy, particularly at positions with more complex multipath environments (e.g., P3, P4, P9). The SS method struggles to provide reliable identification performance in real-world multipath settings. Moreover, the error bars indicate that DeepCRF exhibits relatively consistent performance across different NICs, with a smaller range between the highest and lowest accuracy values compared to other methods. 
It is worth noting that the lowest accuracy values are contributed by identifying devices of the same model. Distinguishing between devices of the same model can be particularly challenging, as they often share similar RF characteristics. However, DeepCRF achieves the highest value of the lowest accuracy among all methods, indicating its superior ability to identify devices of the same model. 
When comparing the accuracy values for $N_{csi}$ equal to 1 and 4, we can observe the improvement brought by the decision fusion method. For all methods, the accuracy values generally go higher as $N_{csi}$ jumps from 1 to 4. Specifically, DeepCRF still achieves the best performance and achieves an average of 99.53\% accuracy of 19 NICs over 7 unseen positions (P3-P9) when $N_{csi}$ is equal to 4.

\subsection{Impact Discussion}
\label{impact}
We now investigate the impact of different design choices on identification accuracy, including fusion methods, input types, and data augmentation methods.

\begin{table}[]
  \caption{Average identification accuracy (\%) of fusion methods.} 
  % \vspace{-0.5em}
  \label{fusion}
  \centering
  \tiny
\resizebox{\linewidth}{!}{
  \begin{tabular}{l|c|c|c|c}
    \toprule
    \multirow{2}{*}{$N_{csi}$} & \multirow{2}{*}{Data}&\multicolumn{3}{c}{Decision Fusion} \\
    \cmidrule{3-5}
    $\times N_{rx}$&Fusion&MV&AP &BC\\
    \midrule
  $1\times1$ &  89.11 & 89.11 & 89.11& 89.11\\ 
  $1\times4$ & 93.35 & 95.90 & \textbf{98.22}& 94.86 \\ 
  $2\times4$ & 95.85 & 98.13 & \textbf{99.14}& 97.10\\ 
  $3\times4$ & 96.80&98.76&\textbf{99.41}&97.78\\
  $4\times4$ & 97.41&99.05&\textbf{99.54}&98.15\\ 
  \bottomrule
\end{tabular}
}

\end{table}

\begin{table}[]
	\caption{Average identification accuracy (\%) of CSI input types: I/Q (Real and Imaginary), A\&P (Amplitude and Phase), and Complex Value ($\mathbb{C}$), when $N_{c} = 1$. }
	% \vspace{-1.5em}
	\label{input}
	\centering
	\subfloat[Synthetic Data]{
		\centering
		\resizebox{\linewidth}{!}{
			\begin{tabular}{l|c|c|c|c|c|c}
				\toprule
				\multirow{2}{*}{Method} & \multicolumn{2}{c|}{B}&\multicolumn{2}{c|}{C} &\multicolumn{2}{c}{D}\\
				\cmidrule{2-7}
				&5-20&25-40&5-20 &25-40 &5-20 &25-40\\
				\midrule
				I/Q& \textbf{32.40}& \textbf{83.57}&29.53& \textbf{82.08}&27.62& 79.14\\
				A\&P& 31.80& 82.77&29.58& 81.34&27.88& 78.69 \\ 
				$\mathbb{C}$ &32.31& 82.81&\textbf{30.01}& 81.13&\textbf{28.79}& \textbf{79.49} \\ 
				\bottomrule
		\end{tabular}}
	}
	
	\subfloat[Real-World Data]{
		\centering
		\resizebox{\linewidth}{!}{
			\begin{tabular}{l|c|c|c|c|c|c|c|c}
				\toprule
				\multirow{1}{*}{Method} & P1-P2 &P3&P4&P5&P6&P7&P8&P9\\
				\midrule
				I/Q&96.91& \textbf{88.81}& \textbf{75.52}& 95.51& 89.68& 89.57& \textbf{97.03}& \textbf{82.25}\\ 
				A\&P&\textbf{97.18}& 88.51& 75.03& 95.28& 89.65& 88.61& 95.33& 82.17\\ 
				$\mathbb{C}$&96.47&88.25&73.96&\textbf{95.58}&\textbf{90.50}& \textbf{89.67}&96.26& 81.73\\ 
				\bottomrule
		\end{tabular}}
	}
	% \vspace{-1em}
\end{table}

\textbf{Fusion Methods}: 
Fusion methods play a vital role in combining the information from multiple CSI measurements to make accurate identifications. In Section~\ref{fusion_method}, we discussed several fusion methods, including data fusion,  Majority Voting (MV), Average Probabilities (AP), and Borda Count (BC). Table~\ref{fusion} shows the average accuracy of 19 NICs over 9 positions for each fusion method, considering different values of the number of CSI instances used for fusion.

As evident from Table~\ref{fusion}, the accuracy of all fusion methods increases as $N_{csi}$ increases. This observation highlights the benefit of leveraging multiple CSI measurements for identification, as it provides more information. However, the performance of data fusion is found to be inferior compared to decision fusion methods. Data fusion involves combining the raw CSI, which are less consistent than learned probabilities. Among the decision fusion methods, the AP method achieves the highest performance. By considering the probabilities rather than just the binary decisions (as in MV) or ranks (as in BC), the AP method can better capture the confidence and uncertainty associated with each identification decision.

\textbf{Input Types}: As mentioned in Section~\ref{netowrk_design}, complex-valued neural networks have the inherent capability to capture the interdependencies between the real and imaginary components of the CSI data. However, they come with the trade-off of requiring twice the amount of parameters compared to real-valued neural networks. In Table~\ref{input}, we present the average accuracy of 19 NICs for each input type, comparing the performance of real-valued and complex-valued neural networks. Our carefully designed real-valued neural network architecture achieves comparable performance to complex-valued neural networks, indicating the efficiency of the proposed design.

Furthermore, Table~\ref{input} shows that using the real and imaginary components of the CSI data as input slightly outperforms using the amplitude and phase components. This difference in performance can be attributed to the distribution characteristics in the dataset. The real and imaginary components of the CSI data are more consistent in terms of their distribution. On the other hand, the amplitude component is always positive (larger than 0), and the phase component is limited to a range from $-\pi$ to $\pi$. These constraints on the amplitude and phase components may introduce biases in the learning process.

\begin{figure}
    \centering
    \includegraphics[width=\linewidth]{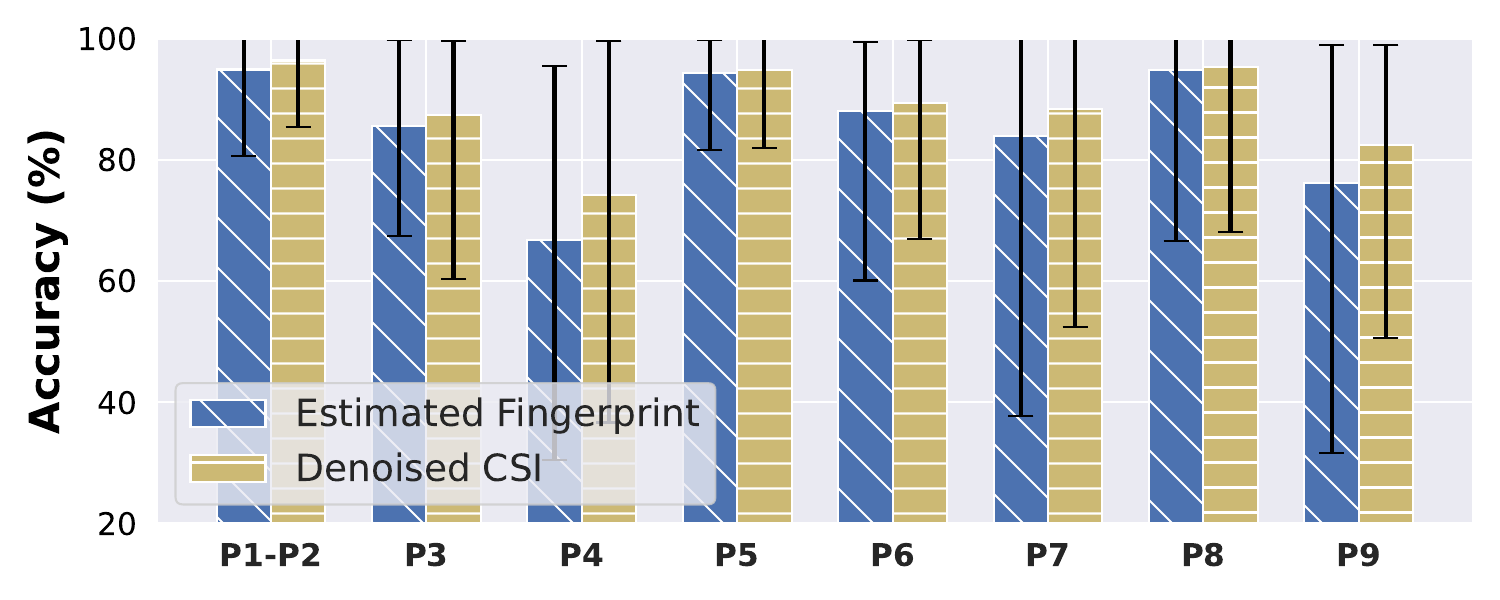}
    \caption{ Performance comparison of different data augmentation methods, when $N_{c} = 1$.}
    \label{aug compare}
    % \vspace{-1em}
\end{figure}

\textbf{Data Augmentation Methods}: In this analysis, we only use the synthetic datasets for training and then test on the real-world dataset to evaluate how closely two different DA methods mimic real-world data. The results presented in Fig.~\ref{aug compare} provide valuable insights into the effectiveness of the two DA methods. Specifically, Fig.~\ref{aug compare} compares the average accuracy among 19 NICs when using estimated fingerprints and denoised CSI for data augmentation. The comparison in Fig.~\ref{aug compare} shows that using denoised CSI for data augmentation consistently outperforms using estimated fingerprints in all scenarios. This finding confirms the discussion in Section~\ref{dataaugmentation} regarding the benefits of integrating the practical filtering effect into the synthetic CSI generation process.

\subsection{Ablations}

\begin{table}[]
	\caption{Results of ablation studies on training strategies: Supervised Contrastive Learning (SCL) and Data Augmentation (DA), when $N_{c} = 1$.}
 % \vspace{-1.5em}
	\label{performance}
	\subfloat[Synthetic Data.]{
		\centering
		\resizebox{\linewidth}{!}{
			\begin{tabular}{l|c|c|c|c|c|c}
				\toprule
				\multirow{2}{*}{Method} & \multicolumn{2}{c|}{B}&\multicolumn{2}{c|}{C} &\multicolumn{2}{c}{D}\\
				\cmidrule{2-7}
				&5-20&25-40&5-20 &25-40 &5-20 &25-40\\
				
				\midrule
				DeepCRF& \textbf{32.40}& \textbf{83.57}&\textbf{29.53}& \textbf{82.08}&\textbf{27.62}& \textbf{79.14}\\
				w/o SCL& 30.65& 82.24&28.24& 81.40&27.36& 78.03\\
				w/o DA   & 13.47& 51.66&11.71& 28.69&10.73& 16.67 \\ 
				w/o DA+SCL  & 13.96& 37.26&12.12& 21.40&10.04& 11.75 \\ 
				\bottomrule
		\end{tabular}}
	}
	\hfill
	\subfloat[Real-World Data]{
		\centering
		\resizebox{\linewidth}{!}{
			\begin{tabular}{l|c|c|c|c|c|c|c|c}
				\toprule
				\multirow{1}{*}{Method} & P1-P2 &P3&P4&P5&P6&P7&P8&P9\\
				\midrule
			DeepCRF&  96.91& \textbf{88.81}& \textbf{75.52}& \textbf{95.51}& 89.68& \textbf{89.57}& \textbf{97.03}& \textbf{82.25}\\ 
				w/o SCL& 97.00&87.64& 72.91& 94.88& \textbf{89.80}& 88.43&94.84&81.20\\ 
				w/o DA   &\textbf{99.91}& 73.34& 54.49& 89.01& 78.65&77.47&85.16&53.35 \\
				w/o DA+SCL  & 99.87& 57.52& 42.11& 71.06& 58.81& 56.28&73.40&32.54 \\ \bottomrule
		\end{tabular}}
	}
	% \vspace{-1em}
\end{table}
The ablation studies presented in Table~\ref{performance} provide compelling evidence for the effectiveness of the training strategies employed in DeepCRF. Table~\ref{performance} presents the average identification accuracy among 19 NICs in both synthetic scenarios and real-world scenarios, highlighting the impact of different training techniques. As Table~\ref{performance} illustrates, incorporating supervised contrastive learning (SCL) into the training regimen yields an average of $1.1\%$ boost in accuracy for the synthetic dataset across all conditions. 
For the real-world dataset, the benefits of SCL are more obvious in environments with more channel variations, yielding improvements of $2.61\%$ and $2.19\%$ at positions P4 and P8, respectively. Such improvement indicates that the supervised contrastive loss used in the initial training stage enhances class separability, which in turn contributes to improved performance. Additionally, it is notable that when the model was trained using a limited dataset of real-world CSI measurements (without data augmentation), the performance gains attributed to SCL were substantial, ranging from 10\% to 20\%. This significant increase underscores the capability of SCL to improve model generalization, especially in scenarios characterized by sparse data.

Besides, the performance of DeepCRF suffers significantly without the implementation of DA, with an average decrease in accuracy of $35\%$ on the synthetic dataset. 
Interestingly, the ablation studies reveal a notable difference in the performance of DeepCRF with and without DA across different positions in the real-world dataset. In positions P1 and P2, the model achieves better performance without DA compared to when DA is applied. However, in positions P3 to P9, the performance of the model without DA is significantly worse than when DA is used. This discrepancy in performance can be attributed to the fact that a portion of the CSI data from positions P1 and P2 is used for training. As a result, the neural network may overfit the specific channel characteristics present in the CSI data from these positions. However, the poor performance of DeepCRF without DA in positions P3 to P9 emphasizes the crucial role of exposing the neural network to CSI data from diverse channel conditions during training. 
Additionally, when DeepCRF is trained without the benefits of both DA and SCL, there is a further decline in performance, as evidenced by an average reduction in accuracy of $40\%$ for the synthetic dataset and $28\%$ for the real-world dataset. This decline underscores the combined contribution of both strategies to the robustness and accuracy of DeepCRF. The more pronounced decline in accuracy on the synthetic dataset can be attributed to the fact that it encompasses a broader spectrum of challenging channel conditions and noise levels than the real-world dataset.

\subsection{Feature Visualization}
\begin{figure}
	\centering    
	\includegraphics[width=0.9\linewidth]{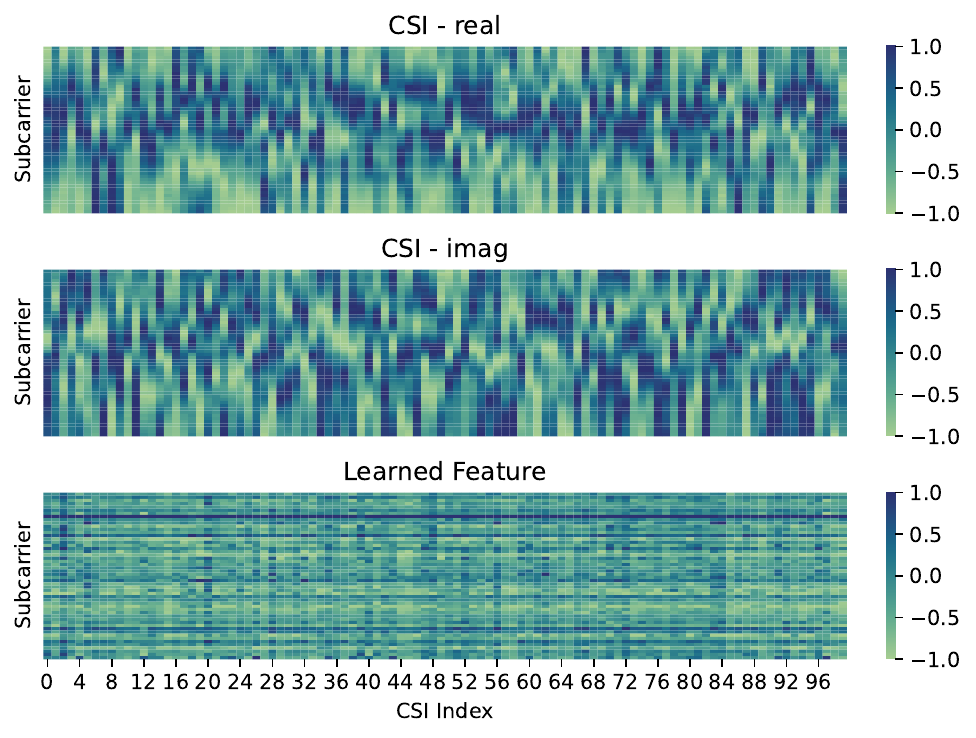}
	\caption{Feature Visualization, under NLoS scenarios of channel Model-D, and SNR = 40dB.}
	\label{visualization}
	% \vspace{-1em}
\end{figure}

We now present an analysis of the fluctuations of CSI and its corresponding learned features across 100 different CSI inputs. Using heatmaps as our visualization tool, we compare the raw CSI samples with the extracted features under NLoS conditions of channel Model-D (i.e., the most complicated multipath scenarios) with a SNR of 40dB. To ensure comparability, both the CSI samples and the learned features are normalized prior to comparison. Fig~\ref{visualization} provides an illustration of the variability inherent in the CSI samples across different channels. In contrast, the learned features through DeepCRF exhibit a remarkable level of stability despite the dramatic changes in the underlying CSI inputs. The consistent pattern observed within the features underscores the efficacy of the feature extraction process and highlights the potential of learned features to accurately capture the RF characteristics of the device from CSI, which are crucial for precise device identification.

\section{Discussions}

\begin{figure}%[h]  
    \centering
  \includegraphics[width=\linewidth]{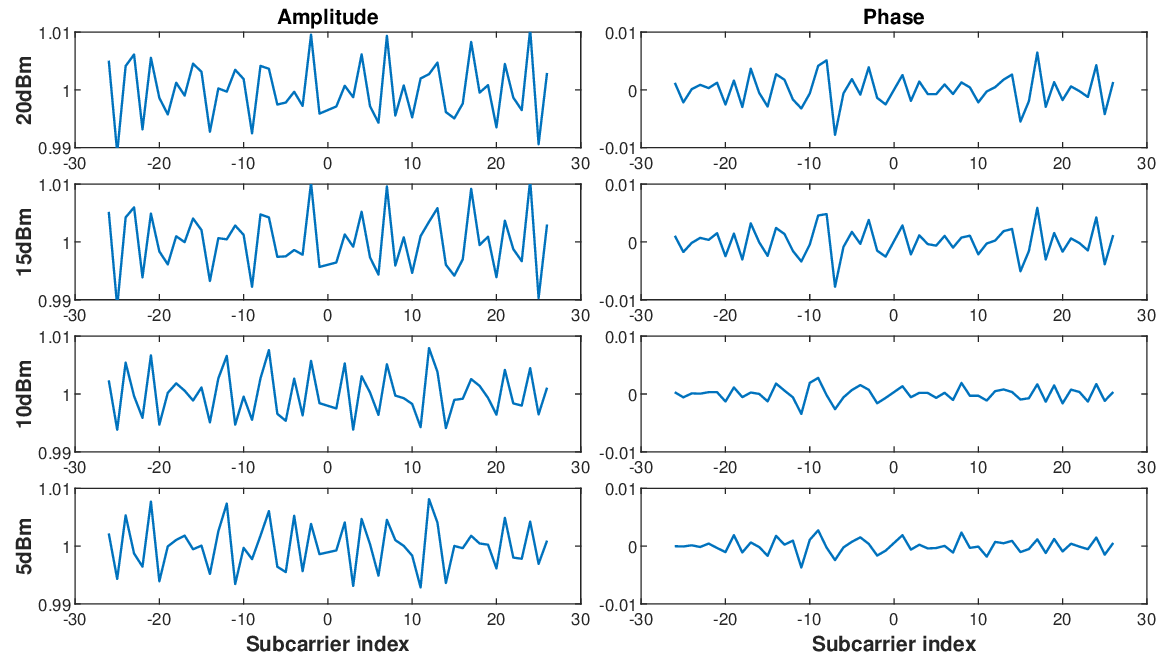}
  \caption{Amplitude and phase of micro-CSI extracted from denoised LoS CSI measurements under different transmit powers. The intensity of micro-CSI changes as the transmit power changes.}
  \label{txpower}
\end{figure}

\subsection{Potential Hardware Cause}
When we conducted experiments to explore the cause of micro-CSI, we found that the intensity of micro-CSI is influenced by transmit power. Specifically, we used an Intel AX200 NIC as the transmitter, and an SDR platform, as described in Section V-A, for receiving. The experiments were conducted in a LoS configuration, with transmit power levels adjusted to 5dBm, 10dBm, 15dBm, and 20dBm via the \texttt{iwconfig} command. The CSI measurements collected at each power level were processed using the SS method \cite{kong2024csirff} to extract the micro-CSI. The findings, depicted in Fig.~\ref{txpower}, clearly show that changes in transmitting power influence the intensity of the micro-CSI. This observation implies that the power amplifier, as a component of the RF transmission chain, likely contributes to the characteristics of micro-CSI.

Our experimental findings that variations in the distance between the transceivers influence the transmitting power, thereby affecting the intensity of micro-CSI. We found that data collected from one LoS and one NLoS scenario were adequate to cover the range of micro-CSI intensities encountered in real-world environments, aiding in more accurate feature learning. 
Comparative analysis reveals a significant change in model accuracy based on the training data used. Specifically, when training exclusively with data from position P1, the average accuracy among P3-P9 decreases from 99.53\% to 94.49\% for $N_{csi} = 4$. This reduction in accuracy can be attributed to the limited range of micro-CSI intensities present in the data from P1. In contrast, incorporating data from both positions P1 and P2 during training provides a more comprehensive range of micro-CSI intensities, leading to improved model performance.

\subsection{Extreme Channel Condition}
To assess performance under rare and extreme conditions, we incorporate synthetic samples based on channels from Model-F during both the training and testing phases. Model-F is characterized by an extreme scenario in which the maximum delay surpasses the cyclic prefix length of the IEEE 802.11 OFDM system, posing significant challenges to communication quality. Despite these harsh conditions, our DeepCRF model demonstrates considerable robustness, achieving an accuracy of 62.98\% in identifying 19 COST devices at an SNR of 40 dB with only a single CSI sample per identification. By comparison, the performance of the other three baselines was much lower, with accuracies of 7.66\% (SS \cite{kong2023physicallayer,kong2024csirff}), 13.73\% (Self-ACC \cite{yang2023eliminating}), and 10.5\% (DeepCSI \cite{meneghello2022deepcsi}) respectively. These results underscore the efficacy of our model in managing severe channel impairments.

\subsection{Incorporation of New Devices}

When incorporating new devices, the initial step involves collecting a small set of CSI measurements from a few positions. Following this, data augmentation is performed as detailed in Section IV-B. The augmented data, along with the newly collected measurements, is then combined with existing training datasets for model retraining. The only modification needed for the DeepCRF model is to adjust the number of neurons in the classifier to match the number of devices. As an example, we conducted experiments incorporating three new devices: MERCURY UX3H, TPLINK XDN6000, and ESP32. Data collection for these devices was done from positions P1 and P2 for training, and from positions P8 and P9 for testing. Retraining the DeepCRF model with 22 classes took approximately 12 minutes using platforms detailed in Section V-C. The results demonstrated high accuracy, achieving 99.98\% on P8 and 99.35\% on P9 when $N_{csi} = 4, N_{rx}=4$. This streamlined process not only demonstrates the efficiency of our data augmentation and neural network design but also confirms the model's generalization capability across an expanded device set.

\subsection{Limitations and Future Works}

The DeepCRF framework, tailored for closed-set identification within a fixed set of devices, needs adaptation to suit the dynamic realities of real-world scenarios where device sets might change. Besides, recent studies have highlighted the vulnerability of DL-based RFFI systems to adversarial attacks. To enhance robustness, we plan to explore several strategies in future work.  Firstly, we will explore incremental learning or few-shot learning techniques, which will enable our system to continually integrate new devices without the need for retraining the entire dataset. This strategy is particularly valuable in dynamic environments, ensuring our system remains both effective and efficient. Secondly, we will conduct systematic evaluations of DeepCRF under adversarial conditions, both in controlled settings and real-world environments, as the dynamic nature of the wireless channel might affect the effectiveness of adversarial attacks. Additionally, we will develop defensive mechanisms specifically designed for the unique properties of RF signals. These will include advanced strategies such as adversarial training and defensive distillation, implemented to protect against adversarial attacks while maintaining robust identification performance under various operating conditions.

\section{Conclusions}

In this paper, we present DeepCRF, a new deep learning-enabled framework tailored for CSI-based radio-frequency fingerprinting (CSI-RFF) of commodity WiFi devices. DeepCRF represents a further step towards channel-resilient CSI-RFF for device identification. By leveraging a thoughtfully designed and trained convolutional neural network complemented by model-inspired data augmentation, DeepCRF has shown a remarkable performance in device identification accuracy, surpassing our prior signal space-based method and state-of-the-art neural network benchmarks. Additionally, DeepCRF incorporates supervised contrastive learning and decision fusion techniques, significantly enhancing its resilience against noise. Our evaluations with synthetic and real-world datasets have confirmed the efficacy of DeepCRF in typical indoor environments under both line-of-sight and non-line-of-sight conditions. Notably, DeepCRF achieves an average of 99.53\% identification accuracy among 19 COST WiFi NICs and requires only 4 CSI measurements per identification procedure.

\bibliography{ref} 
\bibliographystyle{ieeetr}
\end{document}